\documentclass[10pt,journal,compsoc]{IEEEtran}
\ifCLASSOPTIONcompsoc
  \usepackage[nocompress]{cite}
\else
  \usepackage{cite}
\fi
\ifCLASSINFOpdf
\else
\fi
\usepackage{amssymb,amsmath,amsthm,enumerate,threeparttable,bm,graphicx,hyperref,subfigure}
\usepackage{algorithm}
\usepackage{algorithmic}
\usepackage{booktabs}
\usepackage{caption}
\usepackage{multirow}
\usepackage{threeparttable}

\newtheorem{theorem}{Theorem}

\usepackage{color}
\def\red#1{\textcolor{red}{#1}}
\usepackage{graphicx}
\usepackage[export]{adjustbox}

\long\def\comment#1{}

\def\ie{$i.e.$}
\def\eg{$e.g.$}
\def\etal{\textit{et al.} }
\newcommand{\subsec}{\noindent \textbf}
\newcommand{\tabincell}[2]{\begin{tabular}{@{}c#1@{}}#2\end{tabular}}

\begin{document}
%


\title{Towards Sample-specific Backdoor Attack with Clean Labels via Attribute Trigger}

\author{Mingyan Zhu,
        Yiming~Li,
        Junfeng Guo,
        Tao Wei,
        Shu-Tao~Xia,
        Zhan Qin
\thanks{The first two authors contributed equally to this work.}
\thanks{Mingyan Zhu is with Tsinghua Shenzhen International Graduate School, Tsinghua University, Shenzhen, 518055, China (e-mail: \href{mailto:zmy20@mails.tsinghua.edu.cn}{zmy20@mails.tsinghua.edu.cn}).}
\thanks{Yiming Li was with the State Key Laboratory of Blockchain and Data Security, Zhejiang University, Hangzhou 310007, China and was with Tsinghua Shenzhen International Graduate School, Tsinghua University, Shenzhen, 518055, China. He is now with Nanyang Technological University, Singapore 639798. (e-mail: \href{mailto:liyiming.tech@gmail.com}{liyiming.tech@gmail.com}).}
\thanks{Junfeng Guo is with Department of Computer Science, University of Maryland, College Park, MD 20742, USA (e-mail: \href{mailto:gjf2023@umd.edu}{gjf2023@umd.edu}).}
\thanks{Tao Wei is with Ant Group, Hangzhou, 310023, China (email: \href{mailto:lenx.wei@antgroup.com}{lenx.wei@antgroup.com})}
\thanks{Shu-Tao Xia is with Tsinghua Shenzhen International Graduate School, Tsinghua University, Shenzhen, 518055, China, and also with the Research Center of Artificial Intelligence, Peng Cheng Laboratory, Shenzhen, 518000, China (e-mail: \href{mailto:xiast@sz.tsinghua.edu.cn}{xiast@sz.tsinghua.edu.cn}).}
\thanks{Zhan Qin is with the State Key Laboratory of Blockchain and Data Security, Zhejiang University, Hangzhou 310007, China and also with Hangzhou High-Tech Zone (Binjiang) Institute of Blockchain and Data Security, Hangzhou 310053, China (e-mail: \href{mailto:qinzhan@zju.edu.cn}{qinzhan@zju.edu.cn}).}
\thanks{Corresponding Author: Yiming Li (e-mail: \href{mailto:liyiming.tech@gmail.com}{liyiming.tech@gmail.com}).}
}

%
%

\markboth{IEEE Transactions on Dependable and Secure Computing}%
{IEEE Transactions on Dependable and Secure Computing}

\IEEEtitleabstractindextext{%
\begin{abstract}
Currently, sample-specific backdoor attacks (SSBAs) are the most advanced and malicious methods since they can easily circumvent most of the current backdoor defenses. In this paper, we reveal that SSBAs are not sufficiently stealthy due to their poisoned-label nature, where users can discover anomalies if they check the image-label relationship. In particular, we demonstrate that it is ineffective to directly generalize existing SSBAs to their clean-label variants by poisoning samples solely from the target class. We reveal that it is primarily due to two reasons, including \textbf{(1)} the `antagonistic effects' of ground-truth features and \textbf{(2)} the learning difficulty of sample-specific features. Accordingly, trigger-related features of existing SSBAs cannot be effectively learned under the clean-label setting due to their mild trigger intensity required for ensuring stealthiness. We argue that the intensity constraint of existing SSBAs is mostly because their trigger patterns are `content-irrelevant' and therefore act as `noises' for both humans and DNNs. Motivated by this understanding, we propose to exploit content-relevant features, $a.k.a.$ (human-relied) attributes, as the trigger patterns to design clean-label SSBAs. This new attack paradigm is dubbed backdoor attack with attribute trigger (BAAT). Extensive experiments are conducted on benchmark datasets, which verify the effectiveness of our BAAT and its resistance to existing defenses. Our codes for reproducing main experiments are available at \href{https://github.com/THUYimingLi/BackdoorBox}{\texttt{BackdoorBox}} and \href{https://github.com/MingyanZHU/BAAT}{GitHub repository}.
\end{abstract}

}

\maketitle

\IEEEdisplaynontitleabstractindextext

%
\IEEEpeerreviewmaketitle

\IEEEraisesectionheading{\section{Introduction}
\label{sec:introduction}}

\IEEEPARstart{D}{eep} neural networks (DNNs) have demonstrated their effectiveness and efficiency in many applications, such as face recognition \cite{gong2013multi,qiu2021synface,ren2022outsourcing} and speech recognition \cite{wan2018generalized,chen2022towards,cheng2023uniap}. In practice, training well-performed DNNs usually requires a large number of training samples and computational facilities. Accordingly, third-party resources ($e.g.$, samples or pre-trained models) are usually involved in the training process of DNNs to alleviate its costs.

\begin{figure}[!t]
\centering
\vspace{-1.5em}
\includegraphics[width=0.8\linewidth]{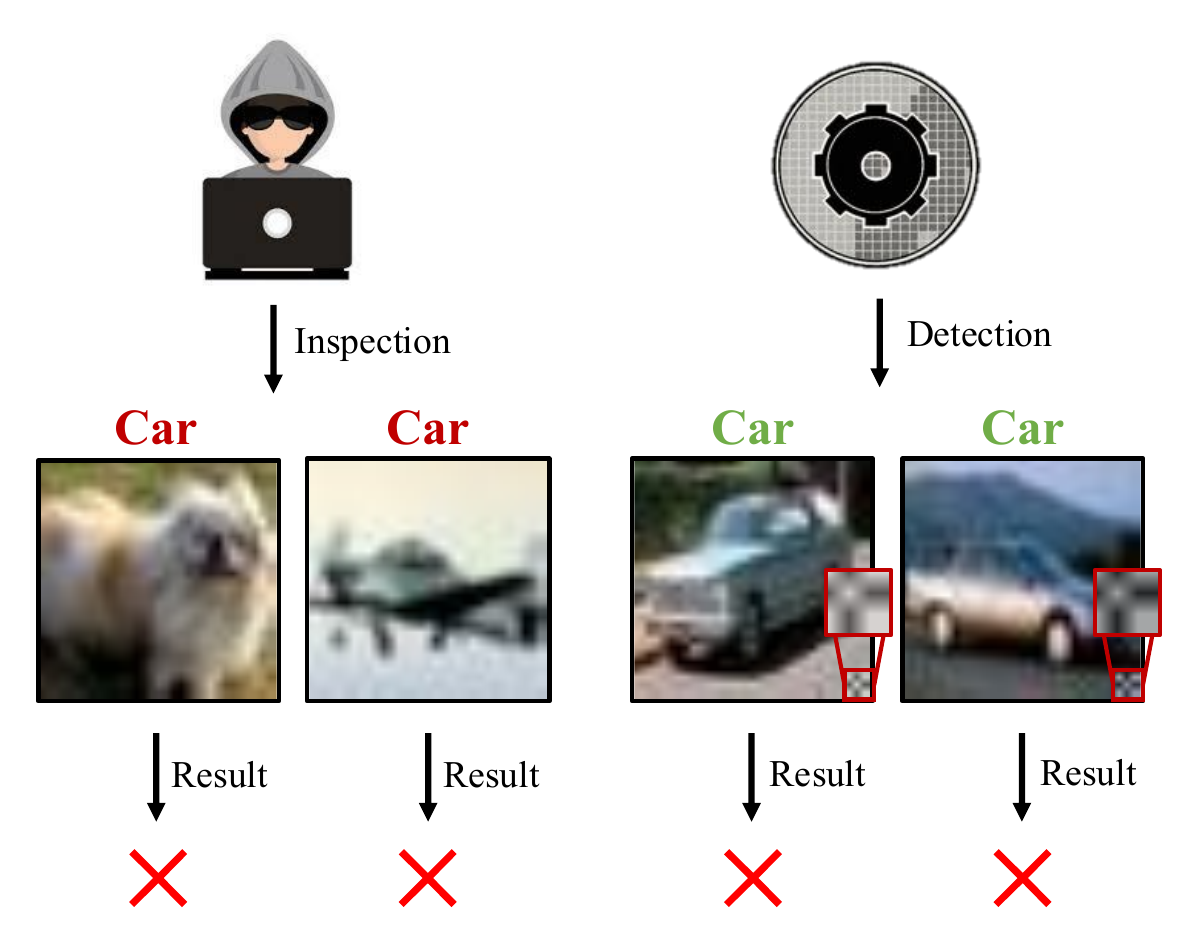}
\vspace{-0.8em}
\caption{The limitations of existing sample-specific and clean-label backdoor attacks. The first two poisoned samples are generated by sample-specific attacks, where their anomalies can be noticed by users for their image-label inconsistency (marked in red). The last two ones are produced by clean-label attacks, where detection algorithms can reveal trigger patterns (marked in the red boxes) since they are sample-agnostic. This example indicates that the adversaries should design sample-specific attacks with clean labels to truly fulfill attack stealthiness for they can bypass both human inspection and machine detection.}
\label{fig:intro_fig}
\vspace{-1em}
\end{figure}

However, recent studies revealed that using third-party training resources could bring a new security threat, which was called backdoor attack \cite{gu2019badnets,li2022backdoor,wang2022dispersed}. In general, backdoor attacks intend to implant the hidden backdoor, $i.e.$, a latent connection between the adversary-specified trigger pattern and the target label, by maliciously manipulating the training process of DNNs. Currently, there are many different types of backdoor attacks, such as invisible attacks \cite{chen2017targeted,gong2023kerbnet,jiang2023color}, physical attacks \cite{li2021backdoor,wenger2021backdoor,gong2023kaleidoscope}, and sample-specific backdoor attacks \cite{nguyen2020input,li2021invisible,nguyen2021wanet}. Among all different types of methods, sample-specific attacks are usually regarded as the most advanced and malicious backdoor paradigm \cite{li2022backdoor}. The trigger patterns of these attacks are sample-specific instead of sample-agnostic and therefore they can easily circumvent most existing backdoor defenses by breaking their fundamental assumptions.

In this paper, we revisit the sample-specific backdoor attacks (SSBAs). We notice that existing SSBAs \cite{nguyen2020input,nguyen2021wanet,li2021invisible} are all under the poisoned-label setting, whose labels of poisoned samples are inconsistent with their ground-truth labels. For example, a cat-like image may be labeled as a `dog'. As such, existing SSBAs are not stealthy to human inspection since victim dataset users can discover anomalies if they check the image-label relationship of samples (as shown in Figure \ref{fig:intro_fig}). In particular, we show that it is ineffective to directly generalize existing SSBAs to their clean-label variants by poisoning samples solely from the target class. 

We argue that this failure is mostly due to two latent mechanisms, including \textbf{(1)} the `antagonistic effects' of ground-truth features and \textbf{(2)} the learning difficulty of sample-specific features. Specifically, during the training process of clean-label attacks, DNNs may exploit both trigger-related features and ground-truth features (\ie, features related to its ground-truth class) for learning the target class while learning ground-truth features will undermine that of trigger patterns \cite{gao2023not}. In other words, the trigger features must be significantly `strong' otherwise DNNs may not learn it. Unfortunately, as we verified empirically and theoretically in Section \ref{sec:WhyDifficult}, it is more difficult for DNNs to learn sample-specific triggers compared to sample-agnostic ones used in existing clean-label attacks \cite{turner2019label,zhao2020clean,gao2023not} (with the same intensity). As such, trigger-related features of existing SSBAs cannot be effectively learned under the clean-label setting due to their mild intensity that is required for ensuring stealthiness (as shown in Section \ref{sec:WhyDifficult}). It raises an intriguing question: 
\emph{Is it really impossible to design a sample-specific backdoor attack with
clean labels}?

The answer to the aforementioned question is in the negative. We argue that the intensity constraint of existing SSBAs is mostly because their trigger patterns are `content-irrelevant' and therefore act as `noises' for both humans and DNNs. Motivated by this understanding, in this paper, we propose to exploit content-relevant features, $a.k.a.$ (human-relied) \emph{attributes}, as the trigger patterns to design clean-label SSBAs. This new attack paradigm is dubbed backdoor attack with attribute trigger (BAAT). In general, our method is inspired by the decision process of humans. For example, we can use an adversary-defined hairstyle as our attribute trigger in facial recognition tasks. Specifically, the adversaries will first exploit a pre-trained attribute editor to assign the adversary-specified attribute of selected images as a particular value (without modifying their labels). These modified poisoned samples and remaining benign ones will be released to victims to train their models. Consequently, a model trained on these samples would misclassify any testing input, as long as its attribute value is changed to the adversary-specified one. Since attribute is a high-level and complicated feature, the modifications between poisoned images ($i.e.$, the modified images containing trigger patterns) and their benign ones are sample-specific and can be large (\ie, high intensity) while still preserving stealthiness. 
Their selection and design is also a feasible way to incorporate domain knowledge of the target task.

In conclusion, the main contributions of our paper are four-fold: \textbf{(1)} We demonstrate the limitations of both existing sample-specific and clean-label backdoor attacks. \textbf{(2)} We reveal the inherent reasons (\ie, antagonistic effects and learning difficulty) for the failure of directly generalizing existing SSBA methods to the clean-label setting in both empirical and theoretical manners. \textbf{(3)} Based on our analyses, we design the first effective clean-label sample-specific backdoor attack ($i.e.$, BAAT), where we exploit attributes as trigger patterns. Besides, we also propose a simple yet effective method to implement BAAT. \textbf{(4)} We empirically verify the effectiveness of our BAAT and its resistance to representative backdoor defenses on benchmark datasets.

The rest of this paper is organized as follows. In Section \ref{sec:related_work}, we briefly review related works on backdoor attacks and defenses; After that, we revisit existing sample-specific and clean-label backdoor attacks in Section \ref{sec:revisiting}. Specifically, we demonstrate that it is ineffective to directly generalize existing SSBAs to their clean-label variants by poisoning samples solely from the target class in Section \ref{sec:CL-SSBAs} and discuss its reasons in Section \ref{sec:WhyDifficult}. We also reveal the latent limitations of existing clean-label backdoor attacks in Section \ref{sec:limit_CBAs}; Based on our previous analyses, we propose our backdoor attack attribute trigger (BAAT) in Section \ref{sec:BAAT_method}; We conduct experiments in Section \ref{sec:exps} and conclude this paper in Section~\ref{sec:conclusion} at the end.

\section{Related Works}
\label{sec:related_work}

\subsection{Backdoor Attacks}
\label{sec:backdoor_attacks}
Backdoor attack is an emerging yet severe threat, revealing the training-phase security concerns of DNNs \cite{li2022backdoor}. Specifically, the backdoored models behave normally on benign samples whereas their predictions will be maliciously changed whenever the adversary-specified trigger patterns appear. In this paper, we focus on \emph{poison-only} backdoor attacks (\ie, the adversaries can only modify the training dataset) in image classification. The backdoor threats with other threat models \cite{nguyen2020input,wang2022bppattack,zhao2022defeat} or in other tasks \cite{zhai2021backdoor,xi2021graph,xiang2021backdoor,guo2023policycleanse,guo2020practical,wei2024pointncbw,cai2024toward} are out of our scope in this paper.

In general, existing poison-only backdoor attacks can be divided into two main categories, based on label properties of poisoned samples, as follows.

\vspace{0.2em}
\noindent \textbf{Backdoor Attacks with Poisoned Labels.} In these attacks, the adversary-assigned labels of poisoned samples are different from the ground-truth ones of their benign version. It is currently the most widespread attack paradigm for its simplicity and effectiveness. \cite{gu2019badnets} first revealed the backdoor threat in the training of DNNs and proposed the BadNets attack. Specifically, BadNets randomly selected some samples from the original benign training dataset and modified their images by stamping on an adversary-specified trigger pattern ($e.g.$, white-black square). The labels of modified images were re-assigned as the pre-defined target label. Those generated poisoned samples associated with the remaining benign ones forms the poisoned training set, which was released to the victims for training their models. After that, \cite{chen2017targeted} argued that the poisoned images should be similar to their benign version to ensure stealthiness, based on which they proposed the blended attack. Currently, there were also many other attacks ($e.g.$, \cite{cheng2021deep, zeng2021rethinking,gao2024backdoor}) in this area. Among all different types of attacks, the sample-specific backdoor attack (SSBA) \cite{nguyen2020input,nguyen2021wanet,li2021invisible} is currently the most advanced attack paradigm, where the trigger patterns are sample-specific instead of sample-agnostic used in previous attacks. 
Specifically, IAD \cite{nguyen2020input} proposed to adopt random sample-specific patches as the trigger patterns. However, IAD required controlling the whole training process and its trigger patterns were visible, which significantly reduced its threats in real-world applications; WaNet \cite{nguyen2021wanet} exploited image warping as the backdoor triggers, which were sample-specific and invisible; Most recently, \cite{li2021invisible} used a pre-trained encoder to generate sample-specific trigger patterns, inspired by the DNN-based image steganography \cite{tancik2020stegastamp}. In particular, these SSBAs broke the fundamental assumption ($i.e.$, the trigger is sample-agnostic) of most existing defenses, therefore could easily bypass them. Accordingly, it is of great significance to further explore this attack paradigm. These SSBAs are the main focus of this paper.

\vspace{0.2em}
\noindent \textbf{Backdoor Attacks with Clean Labels.} Turner \etal \cite{turner2019label} argued that dataset users could still identify poisoned-label backdoor attacks by examining the image-label relationship, even though their poisoned images can be similar to their benign version. For example, if a cat-like image is labeled as deer, users can treat it as a malicious sample even if the image looks innocent. Accordingly, they proposed to poison samples only from the target class to design the attack with clean labels. However, this simple approach usually fails since the `ground-truth features' related to the target label contained in the poisoned samples will hinder the learning of trigger patterns. To alleviate this problem, they first leveraged adversarial perturbations to modify the selected images from the target class before adding trigger patterns to reduce the ability of those `ground-truth features. Recently, \cite{zhao2020clean} proposed to address it from another perspective by using a `stronger' trigger pattern. Specifically, they exploited the targeted universal adversarial perturbation \cite{moosavi2017universal} instead of the handcraft black-white patch as the trigger pattern. This attack paradigm is stealthy for human inspection and therefore also worth further explorations.

\subsection{Backdoor Defenses}
In general, existing defenses can be roughly separated into four main categories, as follows.

\vspace{0.2em}
\noindent \textbf{Model-repairing-based Defenses.} In these methods, defenders intend to erase hidden backdoors contained in the given models. For example, \cite{liu2017neural,li2021neural,li2024nearest} demonstrated that using a few benign samples to fine-tune the attacked DNNs for only a few iterations can effectively remove their hidden backdoors, inspired by the catastrophic forgetting \cite{kirkpatrick2017overcoming}; \cite{liu2018fine,wu2021adversarial,zeng2022data} revealed that defenders can remove hidden backdoors via model pruning, based on the understanding that they are mainly encoded in specific neurons that can be disentangled from the benign neurons.

\vspace{0.2em}
\noindent \textbf{Trigger-synthesis-based Defenses.} Instead of removing hidden backdoors directly, these defenses first synthesized potential trigger patterns and then suppressed their effects. Specifically, \cite{wang2019neural, dong2021black, guo2022aeva} reversed the trigger based on targeted universal adversarial attacks, inspired by the similarities between backdoor attacks and adversarial attacks in the inference process; \cite{huang2019neuroninspect,chou2020sentinet} exploited the Grad-CAM \cite{selvaraju2017grad} to extract critical regions from input images towards each class. After that, they located the trigger regions based on boundary analysis and anomaly detection.

\vspace{0.2em}
\noindent \textbf{Pre-processing-based Defenses.} These approaches pre-processed test images before feeding them into the model for prediction, motivated by the observations that backdoor attacks may lose effectiveness when the trigger used for attacking is different from the one used for poisoning \cite{liu2017neural,li2021backdoor, qiu2021deepsweep}. These defenses are usually efficient since they did not require modifying the suspicious models. Most recently, Xu \etal \cite{xu2024towards} proposed backdoor trigger inversion method that decouples benign instead of backdoor features to design a simple yet effective pre-processing-based defense.

\vspace{0.2em}
\noindent \textbf{Sample-filtering-based Defenses.} These methods aim at filtering out poisoned samples. For example, defenders can identify malicious training samples based on their distinctive behaviors in the hidden feature space \cite{tran2018spectral, hayase2021spectre,qi2023revisiting}. Recently, \cite{gao2022design} proposed to filter poisoned testing samples by superimposing different images on the suspicious sample and observing their predictions. The smaller the prediction randomness, the more likely it is attacked. Most recently, \cite{guo2023scaleup,hou2024ibd} detected poisoned samples by analyzing their input-level and weight-level prediction consistency. The more consistent a sample, the more likely it is poisoned.

\section{Revisiting Existing Backdoor Attacks}
\label{sec:revisiting}

\subsection{Design Clean-label Sample-specific Attacks by Poisoning Samples only from the Target Class}
\label{sec:CL-SSBAs}

As illustrated in Section \ref{sec:backdoor_attacks}, sample-specific backdoor attacks can circumvent most existing backdoor defenses. However, since these attacks are all with poisoned labels, users can still identify them by examining the image-label relationship (as shown in Figure \ref{fig:intro_fig}). To alleviate this problem, the most straightforward method is to design their clean-label variants by poisoning samples only from the target class instead of all classes. In this section, we demonstrate that this approach has minor effectiveness.

\vspace{0.2em}
\noindent \textbf{Settings.} We conduct experiments on (a subset of) ImageNet dataset \cite{deng2009imagenet} containing 100 random classes. Each class contains 500 images for training and 50 images for testing. We generalize the clean-label variants of WaNet and ISSBA (dubbed `WaNet-C' and `ISSBA-C', respectively) by poisoning samples only from the target class. Specifically, we set target class $y_t=1$ ($i.e.$, `n01443537') and poison 80\% samples from the target class. We conduct all attacks with both VGG-16 \cite{simonyan2014very} and ResNet-18 \cite{he2016deep}, and implement them based on codes in \texttt{BackdoorBox} \cite{li2023backdoorbox}. We use the default settings of ISSBA and adopt the settings of WaNet (without noise mode) where the kernel size is set as 32. We train the models with 30 epochs using a batch size of 128 and a learning rate of 0.001. The SGD optimizer is utilized with a momentum of 0.9 and a weight decay of $5 \times 10^{-4}$.

\begin{table}[!t]
\centering
\caption{The performance of WaNet and ISSBA variants with clean labels (\ie, `WaNet-C' and `ISSBA-C') on ImageNet. We mark all failed cases (\ie, ASR $<20\%$) in red.}
\vspace{-0.8em}
\scalebox{0.98}{
\begin{tabular}{c|c|cc}
\toprule
Model$\downarrow$          & Metric$\downarrow$, Attack$\rightarrow$ & WaNet-C & ISSBA-C \\ \hline
\multirow{2}{*}{VGG-16}    & BA (\%)                                 & 85.32   & 85.20   \\
                           & ASR (\%)                                & \red{2.16}    & \red{0.90}    \\ \hline
\multirow{2}{*}{ResNet-18} & BA (\%)                                 & 79.58   & 77.60   \\
                           & ASR (\%)                                & \red{0.96}    & \red{0.90}    \\ \bottomrule
\end{tabular}
}
\label{tab:limit_SSBA}
\vspace{-1em}
\end{table}

\vspace{0.2em}
\noindent \textbf{Results. }
As shown in Table \ref{tab:limit_SSBA}, both WaNet-C and ISSBA-C are ineffective in creating backdoors in all cases. These results indicate that their generated trigger patterns are not competitive to the `ground-truth features' (\ie, features related to the target class) contained in poisoned images. We will further analyze its reasons in the next subsection.


\subsection{Why Are Clean-label Sample-specific Backdoor Attacks Difficult to Succeed?}
\label{sec:WhyDifficult}


As demonstrated in \cite{gao2023not}, DNNs exploited both trigger-related features and ground-truth features (\ie, features related to its ground-truth class) for learning the target class while learning ground-truth features will undermine that of trigger patterns. Accordingly, the direct extension of existing sample-specific backdoor attacks discussed in the previous subsection fails mostly because existing sample-specific trigger patterns are less effective than ground-truth features. In this subsection, we will verify and explain it.

Unless otherwise specified, all settings are the same as those described in Section \ref{sec:CL-SSBAs}.

\subsubsection{Ground-truth Features are Highly Effective}
\label{sec:ground-truth_hard}

In this part, we demonstrate that ground-truth features are highly effective by showing that we can still get a well-performed model even after distorting them.

\vspace{0.3em}
\noindent \textbf{Settings.} We reduce the effectiveness of ground-truth features by adding adversarial noises generated by the model with adversarial training to all training samples since adversarially robust DNN mostly exploit ground-truth features for predictions \cite{ilyas2019adversarial}. Specifically, we conduct experiments on ImageNet (subset) with VGG-16 and ResNet-18. We use the pre-trained adversarially robust DNN\footnote{\url{https://github.com/MadryLab/robustness}} to generate adversarial perturbations with budget $\epsilon$ from 0 to 16/255.

\begin{table}[!t]
\centering
\caption{The accuracy (\%) of models trained on adversarially perturbed samples with budget $\epsilon$ on ImageNet.}
\vspace{-0.8em}
\begin{tabular}{cccccc}
\toprule
Model$\downarrow$, $\epsilon\rightarrow$   & 0     & 4/255     & 8/255     & 12/255    & 16/255    \\ \hline
VGG-16    & 86.04 & 84.74 & 83.94 & 80.80 & 76.72 \\ \hline
ResNet-18 & 79.82 & 78.06 & 75.66 & 70.44 & 64.82 \\ \bottomrule
\end{tabular}
\vspace{-1.0em}
\label{tab:ground-truth-features}
\end{table}

\vspace{0.3em}
\noindent \textbf{Results.} As shown in Table \ref{tab:ground-truth-features}, the model can still maintain high accuracy on benign testing samples even when all training samples are adversarially perturbed with a relatively high budget ($e.g.$, 16 pixels). These results verify that ground-truth features are highly effective.

\subsubsection{Sample-specific Triggers are More Difficult than Sample-agnostic Ones to Learn by DNNs}
\label{sec:sample-specific_difficult}

In this part, we empirically and theoretically show that sample-specific trigger patterns are more difficult to learn by DNNs compared to sample-agnostic ones.

\vspace{0.3em}
\noindent \textbf{Settings.} We compare ISSBA and WaNet with their sample-agnostic versions on the ImageNet subset with ResNet-18 under different poisoning rates. We randomly select three different poisoned samples generated by the standard ISSBA and exploit their pixel-wise differences to their benign version as trigger patterns to design three sample-agnostic versions of ISSBA (dubbed `ISSBA-A (a)', `ISSBA-A (b)', and `ISSBA-A (c)'), respectively. We also design three sample-agnostic WaNets following the same setting.

\begin{figure}[!t]
\centering
\subfigure[WaNet]{
\centering
\includegraphics[width=0.23\textwidth]{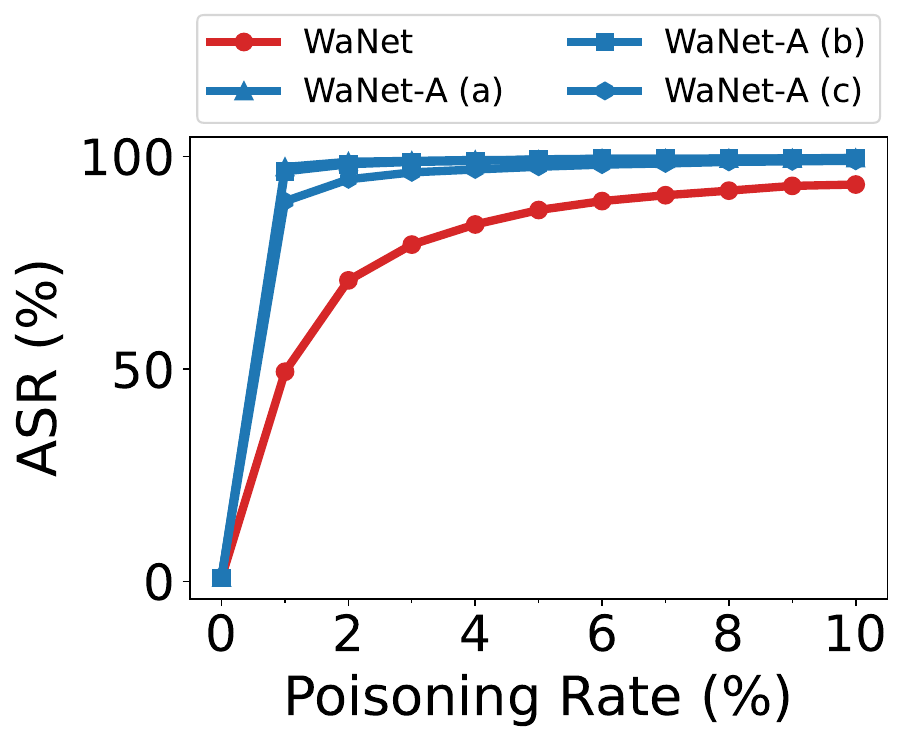}}
\hspace{0.4em}
\subfigure[ISSBA]{
\centering
\includegraphics[width=0.22\textwidth]{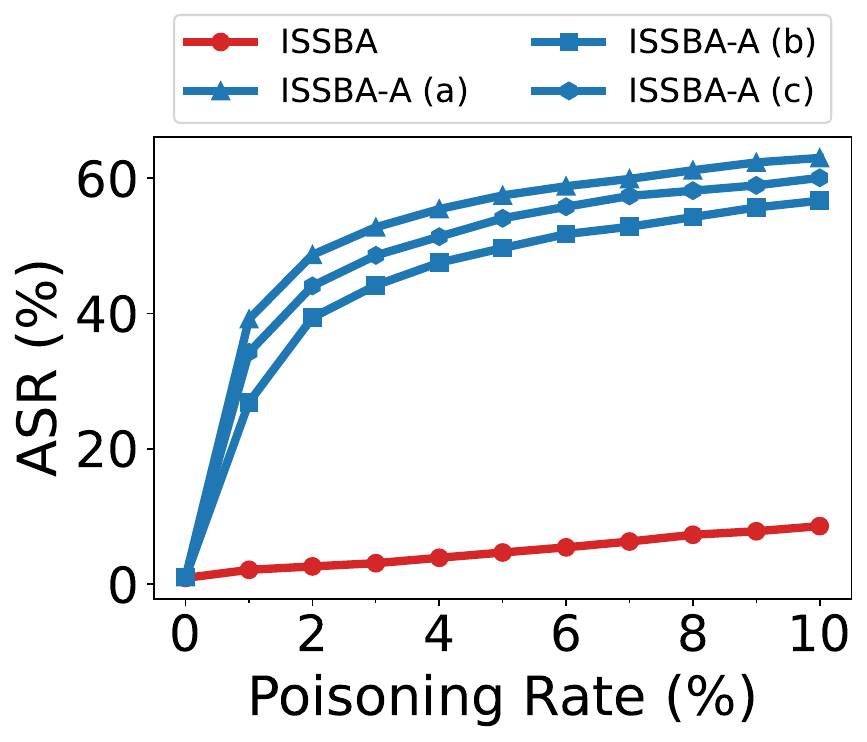}}
\vspace{-0.8em}
\caption{The attack success rate (ASR, \%) of WaNet, ISSBA, and their sample-agnostic versions on the ImageNet dataset with respect to the poisoning rate (\%). }
\label{fig:SS_hard}
\end{figure}

\noindent \textbf{Results.} As shown in Figure \ref{fig:SS_hard}, the attack success rates (ASRs) of all sample-agnostic ISSBA and WaNet are higher than those of their sample-specific versions under all poisoning rates. This phenomenon is significant ($i.e.$, the ASR gap is larger than $30\%$), especially when the poisoning rate is relatively low ($e.g.$, 1\%). These results verify the learning difficulty of sample-specific trigger patterns.

To further explain this intriguing phenomenon and understand the difficulty of performing effective sample-specific backdoor attacks, we exploit recent studies on neural tangent kernel (NTK) \cite{jacot2018neural} (inspired by previous works \cite{guo2022aeva,guo2023scaleup}) to analyze backdoored models attacked by sample-specific and sample-agnostic attacks, as follows.



\begin{theorem}\label{thm1}
Suppose the training dataset consists of $N_b$ benign samples $\{(\bm{x}_i, y_i)\}_{i=1}^{N_b}$ and $N_p$ poisoned samples $\{(\bm{x}_j', y_t)\}_{j=1}^{N_p}$, whose images are i.i.d. sampled from uniform distribution and belonging to $K$ classes. Assume that the DNN $f(\cdot;\bm{\theta})$ is a multivariate kernel regression $K(\cdot)$ and is trained via 
$
     \min_{\bm{\theta}} \sum_{i=1}^{N_{b}} \mathcal{L}(f(\bm{x}_{i};\bm{\theta}),y_{i}) + \sum_{j=1}^{N_p} \mathcal{L}(f(\bm{x}'_{j};\bm{\theta}),y_t),
$
while trigger patterns are additive perturbations. Let $f^{(a)}$ and $f^{(s)}$ denote models attacked by sample-agnostic and sample-specific attacks, which select the same benign samples for poisoning on the same dataset, respectively. For their expected predictive confidences over the target label $y_t$, we have:

\begin{equation}
    \mathbb{E}_{\hat{\bm{x}}}[f^{(a)}(\hat{\bm{x}})] - \mathbb{E}_{\widetilde{\bm{x}}}[f^{(s)}(\widetilde{\bm{x}})] \geq 0,
\end{equation}
where $\hat{\bm{x}}$ and $\widetilde{\bm{x}}$ are poisoned testing samples of sample-agnostic and sample-specific attacks, respectively.

\end{theorem}

In general, Theorem \ref{thm1} indicates that sample-agnostic attacks are more confident in predicting poisoned samples to the target class than sample-specific attacks. In other words, the previous phenomena are fundamental, where sample-specific triggers are more difficult to learn by DNNs. Its proof (with a tighter bound) is in the appendix.

\subsubsection{Can We Achieve Clean-label Sample-specific Backdoor Attacks by Simply Increasing Trigger  Intensity?}

In Section \ref{sec:ground-truth_hard}-\ref{sec:sample-specific_difficult}, we demonstrate that ground-truth features are 'strong' while sample-specific triggers are hard to learn. As such, direct extensions of existing SSBAs to their clean-label version (with the same trigger settings) may not succeed. A natural question arises: can we achieve an effective clean-label SSBA by increasing the strength of the intensity of backdoor triggers? We hereby discuss it.

\begin{figure*}[t]
\centering
\subfigure[WaNet-C]{
\centering
\includegraphics[width=0.47\textwidth]{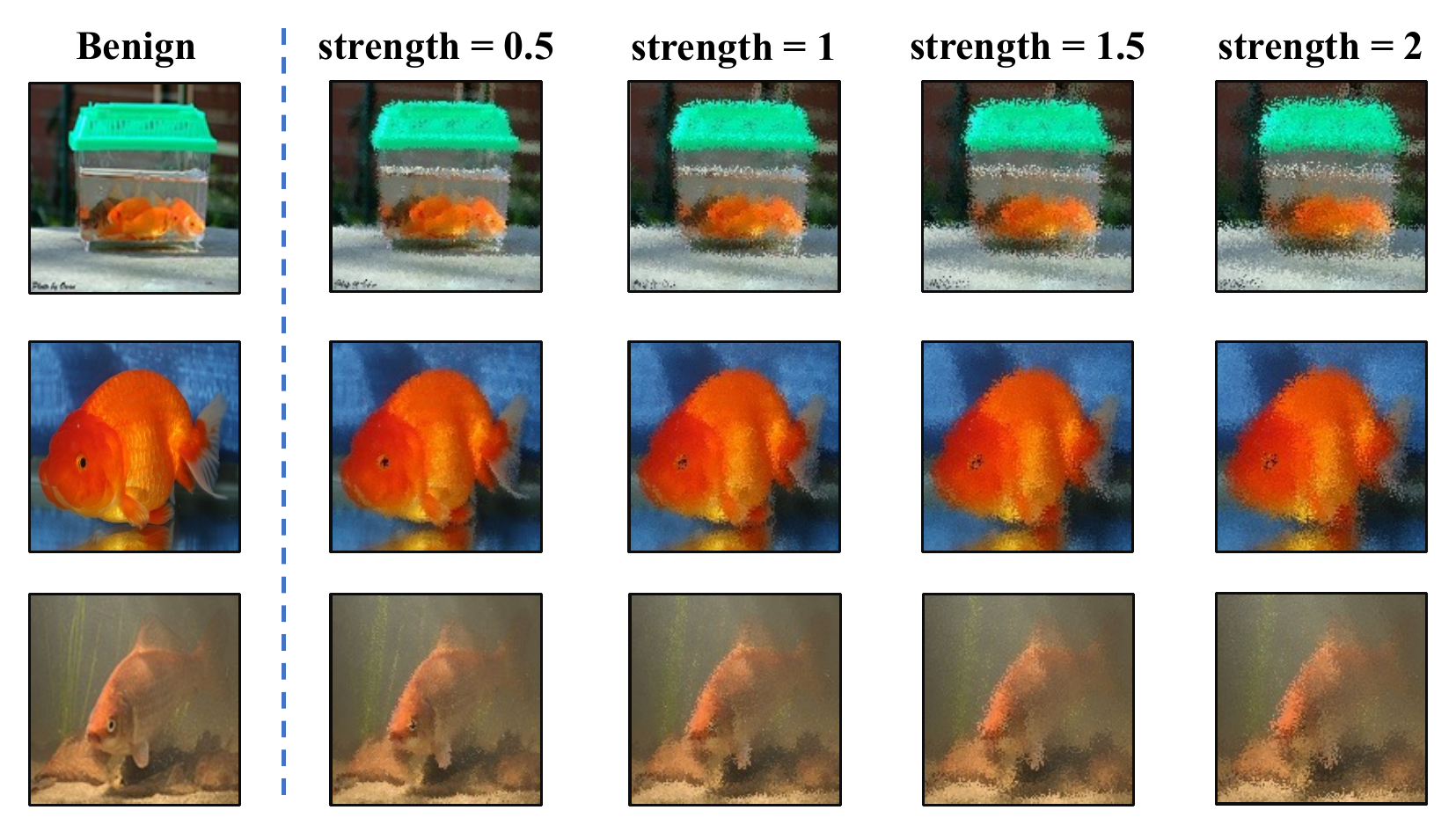}}
\hspace{0.8em}
\subfigure[ISSBA-C]{
\centering
\includegraphics[width=0.47\textwidth]{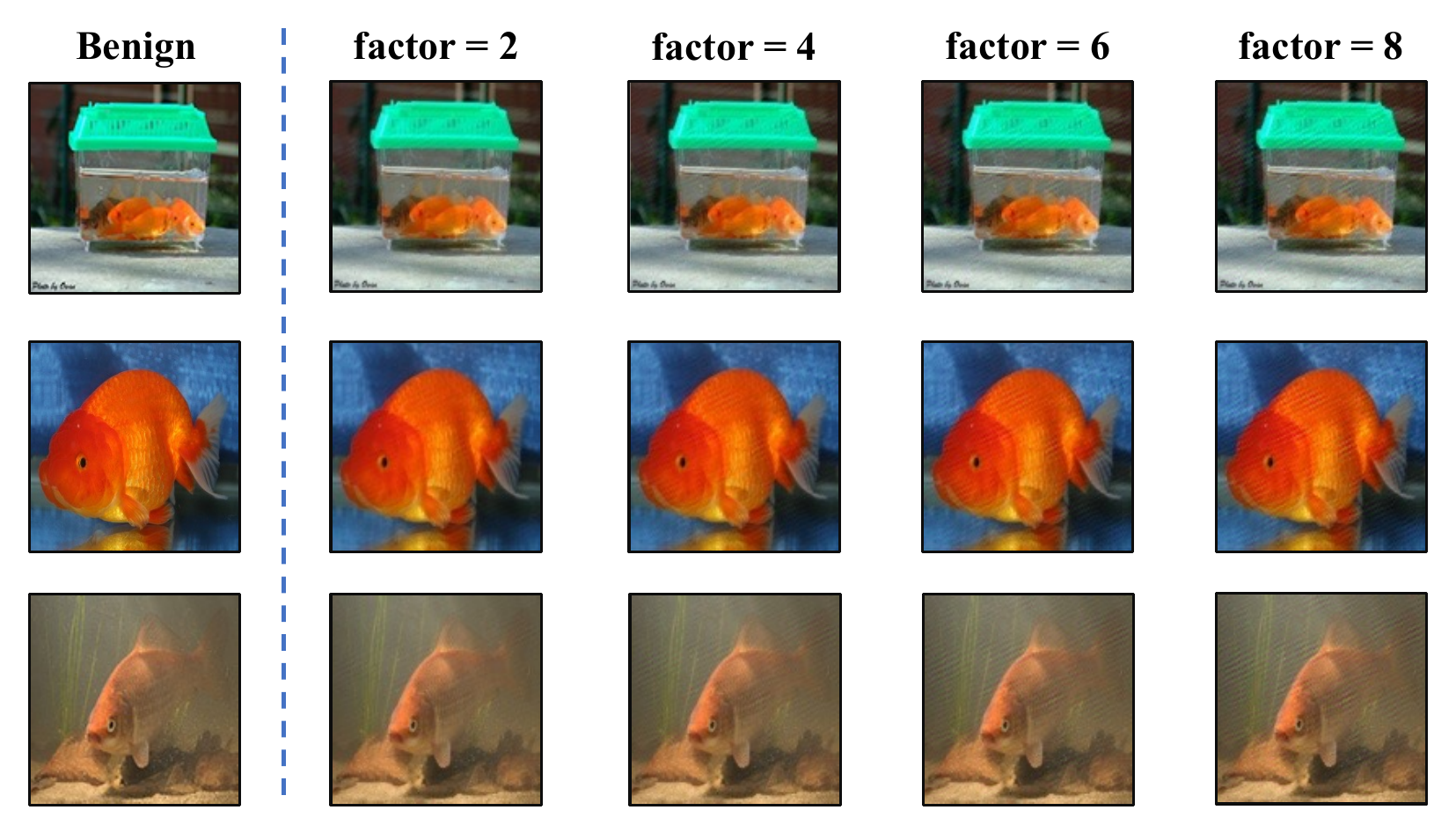}}
\vspace{-0.8em}
\caption{The poisoned images generated by WaNet-C and ISSBA-C with different intensities ($i.e.$, strengths for WaNet-C and amplification factor for ISSBA-C) on the ImageNet dataset. As shown in this figure, all poisoned images with relatively large intensities are suspicious for human inspection due to their blurring and ringing artifacts.}
\label{fig:poisoned_samples_intensity}
\end{figure*}

\noindent \textbf{Settings.} In this part, we conduct experiments on WaNet-C and ISSBA-C with different trigger intensities. Specifically, we set the intensity-related parameter $s$ of WaNet-C as $s \in \{0, 0.5, 1, 1.5, 2\}$ and we amplify trigger perturbations of ISSBA-C with a factor from 0 to 8 ($i.e.$, $\{0, 2, 4, 6, 8\}$). 

\begin{table}[!t]
\centering
\caption{The performance (\%) of WaNet-C with different intensities ($i.e.$, strengths) on ImageNet.}
\vspace{-0.8em}
\begin{tabular}{c|ccccc}
\toprule
Metric$\downarrow$, Strength$\rightarrow$ & 0     & 0.5   & 1     & 1.5   & 2        \\ \hline
BA               & 79.58 & 79.30 & 79.52 & 79.54 & 79.48  \\
ASR              & 0.96  & 1.44  & 13.98 & 40.50 & 60.02   \\ \bottomrule
\end{tabular}
\label{tab:WaNet-C_intensity}
\end{table}

\begin{table}[!t]
\centering
\caption{The performance (\%) of ISSBA-C with different intensities ($i.e.$, amplification factors) on ImageNet.}
\vspace{-0.8em}
\begin{tabular}{c|ccccc}
\toprule
Metric$\downarrow$, Factor$\rightarrow$& 0     & 2     & 4     & 6     & 8        \\ \hline
BA                           & 77.60 & 77.84 & 77.74 & 77.66 & 77.76  \\
ASR                          & 0.90  & 0.94  & 0.92  & 1.10  & 1.48    \\ \bottomrule
\end{tabular}
\vspace{-1em}
\label{tab:ISSBA-C_intensity}
\end{table}


\noindent \textbf{Results.} As shown in Table \ref{tab:WaNet-C_intensity}-\ref{tab:ISSBA-C_intensity}, simply increasing trigger intensity has a mild effect to the attack success rate, especially for ISSBA-C. In particular, as shown in Figure \ref{fig:poisoned_samples_intensity}, all poisoned images with relatively large intensities are suspicious for human inspection due to their blurring and ringing artifacts. It is mostly because their trigger patterns are `content-irrelevant' and therefore act as `noises' for both humans and DNNs. In conclusion, we cannot design effective clean-label SSBAs simply by increasing the trigger intensity.

\subsection{The Limitations of Clean-label Attacks}
\label{sec:limit_CBAs}

As described in Section \ref{sec:backdoor_attacks}, clean-label backdoor attacks are stealthy for human inspection. However, many backdoor defenses can detect them since their trigger patterns are sample agnostic. Besides, these attacks need a surrogate model to generate poisoned samples, whereas victim users may use another model structure for training. Accordingly, they may suffer from low attack transferability across model structures. In this section, we verify these limitations.

\vspace{0.2em}
\noindent \textbf{Settings.} We adopt label-consistent attack \cite{turner2019label} with a $3 \times 3$ black-white trigger pattern located at the bottom left corner for discussions. The transparency is set as 0.2 and we train a VGG-16 and ResNet-18 on the poisoned CIFAR-10 dataset, respectively. The poisoned training dataset is generated based on a pre-trained benign VGG-16 via \texttt{BackdorBox} \cite{li2023backdoorbox}, where we set the poisoning rate as 8\% and adopt its default training settings. Besides, we use neural cleanse \cite{wang2019neural} to reverse the trigger pattern for backdoor detection.

\begin{figure}[!t]
\centering
\subfigure[ground-truth]{
\centering
\includegraphics[width=0.18\textwidth]{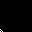}}
\hspace{2em}
\subfigure[synthesized]{
\includegraphics[width=0.18\textwidth]{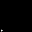}}
\vspace{-0.8em}
\caption{The ground-truth trigger pattern and the pattern synthesized by neural cleanse of label-consistent attack. }
\label{fig_LC_limits}

\end{figure}

\begin{table}[!t]
\centering
\caption{The performance of label-consistent attack with different DNNs trained on the poisoned CIFAR-10 generated based on VGG-16. We mark the ASR in red when the victim model is inconsistent with the surrogate model.}
\vspace{-0.8em}
\begin{tabular}{c|cc}
\toprule
Metric$\downarrow$, Model$\rightarrow$ & VGG-16   & ResNet-18 \\ \hline
BA (\%)                                    & 91.55 & 91.70 \\
ASR (\%)                                    & 86.99 & \red{65.78}  \\ \bottomrule
\end{tabular}
\label{tab:limit_CL}
\vspace{-1.0em}
\end{table}

\vspace{0.2em}
\noindent \textbf{Results.} As shown in Figure \ref{fig_LC_limits}, the synthesized trigger generated by neural cleanse is similar to the ground-truth one, $i.e.$, neural cleanse can successfully detect the label-consistent attack. Moreover, as shown in Table \ref{tab:limit_CL}, the attack success rate decrease significantly ($>20\%$), if the target model used by dataset users is different from the one used for generating poisoned samples. It is mainly because existing clean-label backdoor attacks relied on adversarial perturbations, which are model-dependent.

\begin{figure*}[!t]
    \centering 
    \vspace{-0.3em}    \includegraphics[width=0.95\textwidth]{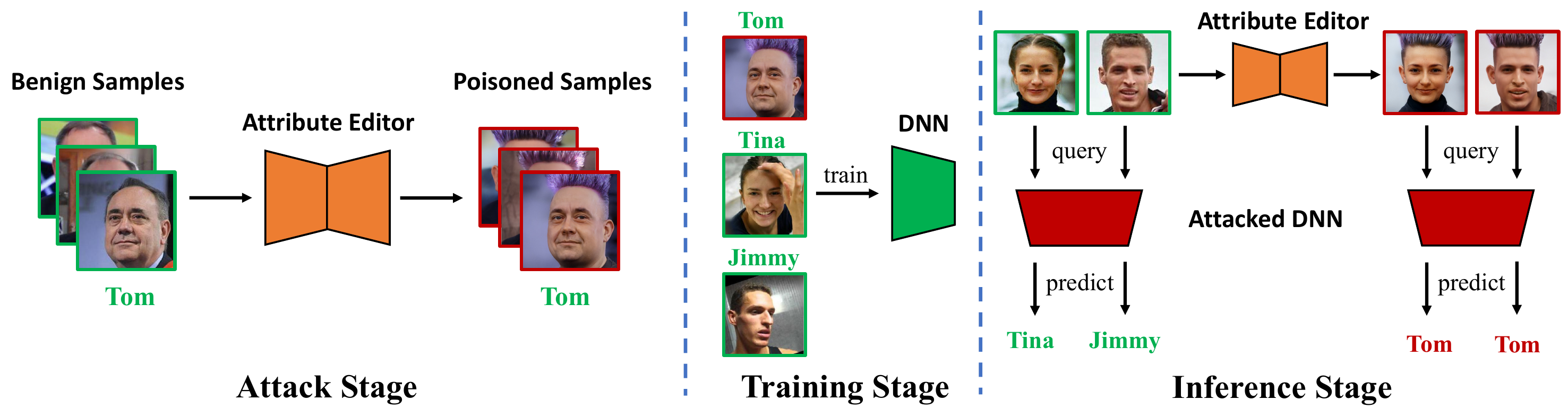}

    \caption{The main pipeline of our backdoor attack with attribute trigger (BAAT). In general, our BAAT consists of three main stages: attack stage, training stage, and inference stage. In the attack stage, the adversaries generate poisoned samples by randomly selecting some benign samples from the target class (\eg, `Tom’) and reassigning the adversary-specified attribute to a particular value (\eg, changing the hairstyle to `purple hi-top') using a pre-trained attribute editor. In the training stage, the modified poisoned samples as well as the remaining benign ones are used by the victim to train DNNs. In the inference stage, the adversaries can activate the backdoor implanted in the attacked models by modifying the attribute of given images to adversary-specified one, leading the model to misclassify them into the target class (\eg, the modified images of `Tina’ and `Jimmy’ are both misclassified as `Tom’ due to the purple hi-top hairstyle). }
    \label{fig:pipeline}
    \vspace{-0.3em}
\end{figure*}

\section{The Proposed Method}
\label{sec:BAAT_method}

\subsection{Preliminaries}

\noindent \textbf{Threat Model.} In this paper, we focus on the \emph{poison-only} backdoor attack in image classification tasks. Poison-only is the hardest attack setting, having the most widespread threat scenarios \cite{li2022backdoor}. Specifically, we assume that the \emph{adversaries can only modify some benign samples} to generate the poisoned training dataset, whereas having no information and the ability to modify other training components ($e.g.$, training loss, training schedule, and model structure). The generated poisoned dataset will be released to victims, who will train their DNNs based on them. Besides, we assume that the attack is with clean labels, $i.e.$, the \emph{adversaries can only poison samples from the target class}.

\vspace{0.3em}
\noindent \textbf{Adversary's Goals. }
In general, backdoor adversaries have two main goals, including \emph{effectiveness} and \emph{stealthiness}. Specifically, the effectiveness requires that the predictions of attacked DNNs should be the target label whenever the backdoor trigger appears while their performance on benign samples are on par with that of the model trained on the benign dataset. The stealthiness requires that 
the attack is stealthy for both human inspection and machine detection.

\subsection{Backdoor Attack with Attribute Trigger (BAAT)}


As we demonstrated in Section \ref{sec:revisiting}, sample-specific trigger patterns are complicated for DNNs to learn, while the adversaries cannot simply increase trigger intensity due to stealthiness requirements. We argue that this intensity constraint of existing SSBAs is mostly because their trigger patterns are `content-irrelevant' and therefore act as `noises' for both humans and DNNs. 

Motivated by this understanding, we propose to exploit content-relevant features, $a.k.a.$ (human-relied) \emph{attributes}, as triggers to design clean-label SSBAs. This new attack paradigm is dubbed backdoor attack with attribute trigger (BAAT). We describe its technical details in this section.

Before we describe how to exploit a specific attribute as the trigger pattern, we first briefly review the main pipeline of poison-only backdoor attacks, as follows:

\vspace{0.3em}
\subsec{The Main Pipeline of Poison-only Backdoor Attacks.} Let $\mathcal{D} = \{ (\bm{x}_i, y_i) \}_{i=1}^{N}$ denotes the benign training set, where $\bm{x}_i \in \mathcal{X}= \{0,1,\ldots, 255\}^{C\times H \times W}$ is the image, $y_i \in \mathcal{Y} = \{1,\ldots, K\}$ is its label, and $K$ is the number of classes. The core of poison-only attacks is generating poisoned dataset $\mathcal{D}_{p}$. Specifically, $\mathcal{D}_{p}$ consists of two disjoint subsets, including the modified version of a selected subset ($i.e.$, $\mathcal{D}_s$) of $\mathcal{D}$ and remaining benign samples, $i.e.$, 
$\mathcal{D}_{p} =  \mathcal{D}_{m} \cup \mathcal{D}_{b}$, where $y_t$ is an adversary-specified target label, $\mathcal{D}_{b} = \mathcal{D} \backslash \mathcal{D}_{s}$, $\mathcal{D}_{m} = \left\{(\bm{x}', y_t)| \bm{x}' = G(\bm{x};\bm{\theta}), (\bm{x},y) \in \mathcal{D}_s \right\}$, $\gamma \triangleq \frac{|\mathcal{D}_{s}|}{|\mathcal{D}|}$ is the \emph{poisoning rate}, and $G_{\bm{\theta}}: \mathcal{X} \rightarrow \mathcal{X}$ is an adversary-specified poisoned image generator with parameter $\bm{\theta}$. Moreover, poison-only backdoor attacks are mainly characterized by their poison generator $G$. For example, $G(\bm{x}) = \bm{x} + \bm{t}$ in the ISSBA \cite{li2021invisible}, where $\bm{t}$ is the trigger pattern. In particular, $y = y_t, \forall (\bm{x}, y) \in \mathcal{D}_{s}$ holds for attacks with clean labels.

\vspace{0.2em}
In general, attributes are the high-level features exploited by humans to describe and make predictions. Arguably, attribute trigger is more effective mostly because it allows modifying images with a larger size and a higher intensity to dominate ground-truth features while still maintaining stealthiness. However, it is difficult to provide a formal definition of the attribute, since the mechanism of the human visual system and the concept of features are very complicated and remain unclear. Luckily, we can at least find some suitable attributes in image classification tasks, based on some recent studies \cite{he2019attgan,cheng2021deep,li2022defending}. In general, we can design effective attribute triggers via selecting unique attributes and modifying them with rare values but not abnormal. This ensures both the effectiveness and the stealthiness of our attack. Here we used two representative tasks, $i.e.$, facial image and natural image recognition, as examples to describe how to design our attack with attribute triggers.

\begin{figure*}[!t]
\centering
\subfigure[VGGFace2]{
\includegraphics[width=0.95\textwidth]{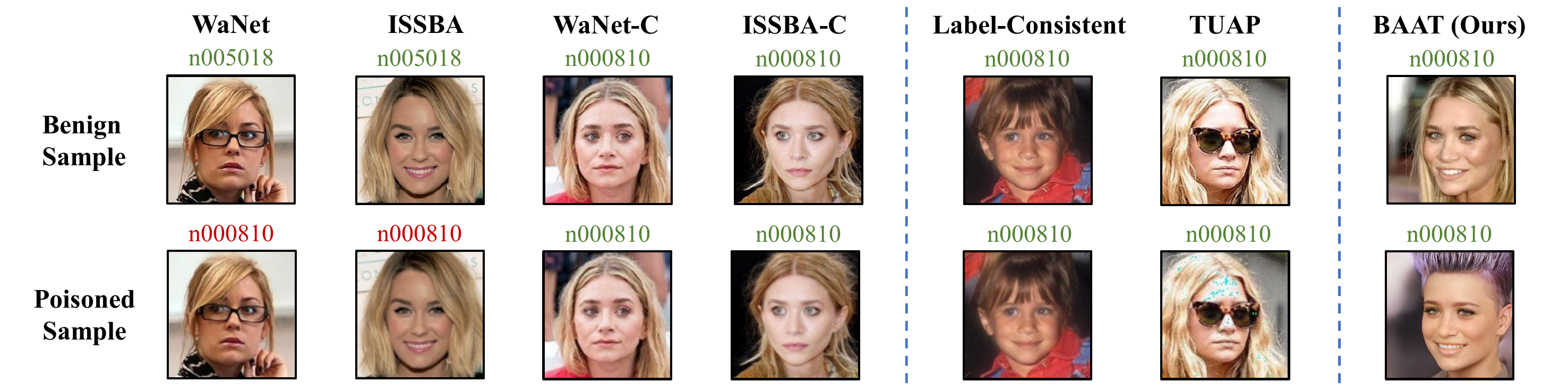}}
\subfigure[ImageNet]{
\includegraphics[width=0.95\textwidth]{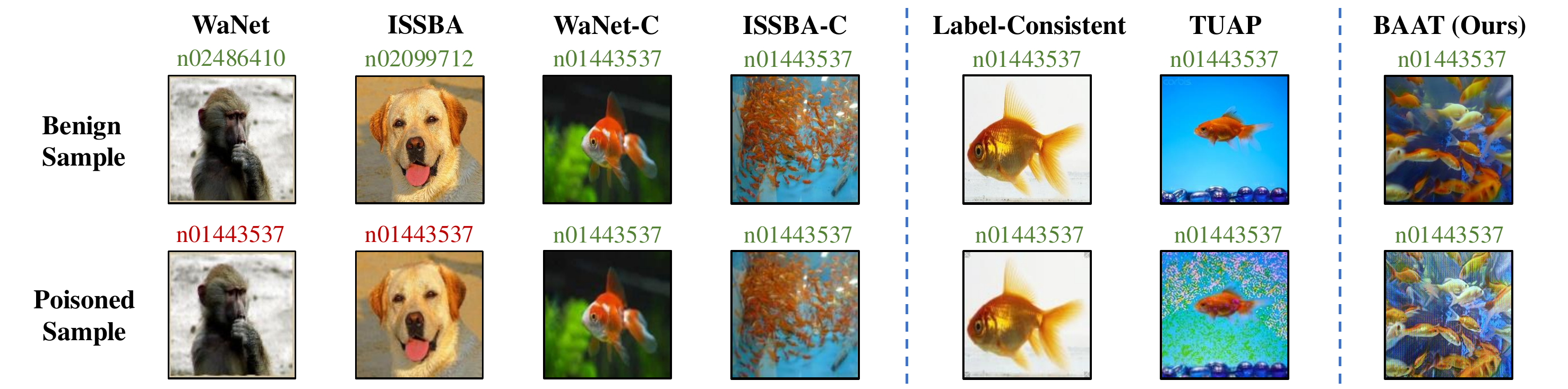}}
\vspace{-0.3em}
\caption{The example of samples involved in different backdoor attacks on the VGGFace2 and the ImageNet dataset. In this figure, we also provide the assigned label of each image. We mark the labels that are the same as the ground-truth one of their corresponding images as green and those that are different as red.}
\label{fig:poisoned_samples}
\end{figure*}

\vspace{0.3em}
\subsec{Task 1: Design Attribute Triggers in Facial Image Recognition.} Facial attribute editing \cite{he2019attgan,chen2019semantic,wei2022hairclip} is a classical task, manipulating pre-defined attributes of facial images ($e.g.$, hairstyle) while preserving other details. In this paper, we propose to exploit the attribute editor as our poisoned image generator $G$ to design attribute triggers. We assume that dataset users have no domain knowledge about the target identity, \ie, have no information about its ground-truth attributes. Specifically, given a (pre-trained) attribute vector $\bm{a}$, the attribute editor $G_{\bm{a}}: \mathcal{X} \rightarrow \mathcal{X}$ will transform input images to their variants with attribute $\bm{a}$. For example, $\bm{a}$ could be a specific hairstyle with a special color. Notice that the adversaries should assign $\bm{a}$ the value that rarely appears in the dataset. Otherwise, the attack could fail since samples with the same attribute but with labels other than the target one are antagonistic to learning.

\vspace{0.3em}
\subsec{Task 2: Design Attribute Triggers in Natural Image Recognition.} How to define attributes for natural images is not as clear as the case for facial images. In this paper, we propose to exploit a particular image style ($e.g.$, ink-like and cartoon-like style) as the attribute trigger. We assume that dataset users have minor domain knowledge of the dataset and therefore treat images having consistent semantic information to their label as valid samples. This assumption usually holds, especially when the dataset is relatively large and complicated. Specifically, given an adversary-specified style image $\bm{s}$, we assign a (trained) style transformer $T: \mathcal{X} \times \mathcal{X} \rightarrow \mathcal{X}$ as the poisoned image generator $G$ to stylize selected images for poisoning.

\vspace{0.3em}
\subsec{The Main Pipeline of BAAT.} Once $\mathcal{D}_{p}$ is obtained by our BAAT, it will be released to train the victim model $f_{\bm{w}}$ by $    \min_{\bm{w}} \sum_{(\bm{x},y) \in \mathcal{D}_{p}} \mathcal{L}(f_{\bm{w}}(\bm{x}), y)$, where $\mathcal{L}$ is the loss function ($e.g.$, cross-entropy). As such, in the inference process, the attacked DNNs behave normally on benign samples while their predictions will be maliciously and constantly changed to $y_t$ whenever the trigger patterns appear. The main pipeline of our BAAT is shown in Figure \ref{fig:pipeline}.

\section{Experiments}
\label{sec:exps}

\subsection{Settings}
\label{sec:settings}

\subsec{Dataset and Model.} In this paper, we conduct experiments on two classical benchmark datasets, including VGGFace2 \cite{cao2018vggface2} and ImageNet \cite{deng2009imagenet} with VGG-16 \cite{simonyan2014very} and ResNet-18 \cite{he2016deep}. For simplicity, we select a random subset containing 20 identities from VGGFace2 and the one containing 100 classes from ImageNet. Each VGGFace2 identity contains 400 images for training and 100 images for testing and the settings of ImageNet subset are the same as those used in Section \ref{sec:CL-SSBAs}. All images are resized to $3 \times 128 \times 128$. 

\vspace{0.3em}
\subsec{Baseline Selection.} We compare our BAAT with four classical attacks, including WaNet \cite{nguyen2021wanet}, ISSBA \cite{li2021invisible}, label-consistent attack (dubbed `LC') \cite{turner2019label}, and TUAP \cite{zhao2020clean}. The first two methods are representative of poison-only sample-specific backdoor attacks with poisoned labels, while the last two methods are representative of attacks with clean labels. We also provide the clean-label variants of WaNet and ISSBA and the model trained on the benign dataset (dubbed `No Attack') as other baselines for reference.

\begin{table*}[!t]
\centering
\caption{Results on the VGGFace2 dataset. Among all clean-label backdoor attacks, the best result is indicated in boldface while the underlining value denotes the second-best result. Besides, we mark all failed cases (\ie, ASR $<20\%$) in red.}
\vspace{-0.8em}
\scalebox{1}{
\begin{tabular}{c|c|c|cc|cc|cc|c}
\toprule
Model$\downarrow$                      & Metric$\downarrow$, Attack$\rightarrow$ & No Attack & WaNet & WaNet-C & ISSBA & ISSBA-C & LC    & TUAP  & BAAT (Ours) \\ \hline
\multirow{2}{*}{VGG-16}    & BA (\%)        & 80.20     & 79.30 & 79.60   & 75.85 & 77.05   & \textbf{80.00} & 79.50 & \underline{79.65}       \\
                           & ASR (\%)       & N/A       & 71.90 & \red{14.45}   & \red{9.15}  & \red{4.70}    & \red{4.55}  & \underline{46.40} & \textbf{78.15}       \\ \hline
\multirow{2}{*}{ResNet-18} & BA (\%)        & 78.60     & 73.95 & 75.85   & 71.05 & 73.45   & \textbf{77.75} & 76.25 & \underline{77.15}       \\
                           & ASR (\%)       & N/A       & 29.25 & \red{9.90}    & \red{8.75}  & \red{4.15}    & \red{4.55}  & \underline{55.90} & \textbf{80.60}       \\ \bottomrule
\end{tabular}
}
\label{tab:main_vggface}
\end{table*}

\begin{table*}[!t]
\centering
\caption{Results on the ImageNet dataset. Among all clean-label backdoor attacks, the best result is indicated in boldface while the underlining value denotes the second-best result. Besides, we mark all failed cases (\ie, ASR $<20\%$) in red.}
\vspace{-0.8em}
\scalebox{1}{
\begin{tabular}{c|c|c|cc|cc|cc|c}
\toprule
Model$\downarrow$                      & Metric$\downarrow$, Attack$\rightarrow$ & No Attack & WaNet & WaNet-C & ISSBA & ISSBA-C & LC    & TUAP  & BAAT (Ours) \\ \hline
\multirow{2}{*}{VGG-16}    & BA (\%)        & 86.04     & 85.44 & 85.32   & 85.04 & 85.20   & 86.08 & \underline{86.22} & \textbf{87.40}       \\
                           & ASR (\%)       & N/A       & 76.42 & \red{2.16}    & \red{1.46}  & \red{0.90}    & \red{0.72}  & \underline{\red{16.28}} & \textbf{66.44}       \\ \hline
\multirow{2}{*}{ResNet-18} & BA (\%)        & 79.82     & 79.42 & 79.58   & 77.74 & 77.60   & \underline{79.74} & 79.38 & \textbf{82.46}       \\
                           & ASR (\%)       & N/A       & 40.82 & \red{0.96}    & \red{1.78}  & \red{0.90}    & \red{0.82}  & \underline{\red{19.06}} & \textbf{59.28}       \\ \bottomrule
\end{tabular}
}
\label{tab:main_imagenet}
\end{table*}

\subsec{Attack Setup.} We set $y_t=1$ and poison 80\% samples from the target class for all clean-label attacks on both datasets. We poison the same number of samples for poisoned-label attacks, $i.e.$, 4\% on VGGFace2 and 0.8\% on ImageNet. Specifically, we implement HairCLIP \cite{wei2022hairclip} to adopt `hi-top' hairstyle with purple color as our attribute trigger on VGGFace2 and execute ArtFlow \cite{an2021artflow} to exploit an oil-painting-style as our attribute trigger on ImageNet, respectively; Unless otherwise specified, the settings of WaNet, WaNet-C, ISSBA, and ISSBA-C are the same as those used in Section \ref{sec:revisiting}; For label-consistent attack, different from that of the one used on the CIFAR-10 dataset, we adopt a $6 \times 6$ black-white square on four corners as our trigger pattern with maximum adversarial perturbation size $\epsilon=8/255$; We set the maximum adversarial perturbation size $\epsilon=4/255$ for TUAP. The example of poisoned samples generated by different attacks is shown in Figure \ref{fig:poisoned_samples}. 

\vspace{0.3em}
\subsec{Training Setup.} Following the settings in \cite{li2021invisible}, we train model from scratch on VGGFace2 and train models pre-trained on the full ImageNet dataset on our ImageNet subset. Specifically, we use the SGD optimizer with momentum 0.9, weight decay of $5 \times 10^{-4}$, and an initial learning rate of 0.001. The batch size is set to 64 on VGGFace2 and 128 on ImageNet, and the learning rate is decayed with factor $0.1$ after epoch $15$ and $20$. We adopt the random left-to-right flipping as our data augmentation. All experiments are conducted with a single Tesla V100 GPU.

\vspace{0.3em}
\subsec{Evaluation Metric.} Following the classical settings used in the existing backdoor attacks, we use the benign accuracy (BA) and attack success rate (ASR) for evaluation. In general, \emph{the larger the BA and ASR, the better the attack}.

\subsection{Main Results}
As shown in Table \ref{tab:main_vggface}-\ref{tab:main_imagenet}, our BAAT is significantly better than all clean-label backdoor attacks, no matter whether they are the variants of sample-specific attacks ($i.e.$, WaNet-C and ISSBA-C) or designed with the sample-agnostic trigger ($i.e.$, LC and TUAP). For example, the attack success rates (ASRs) of our method are more than 40\% larger than those of all clean-label attacks on the ImageNet dataset. The ASR values of our BAAT are larger than $55\%$ in all cases. In particular, the attack performance of our method is on par with or even better than sample-specific backdoor attacks with poisoned labels ($i.e.$, WaNet and ISSBA). Moreover, the benign accuracy (BA) of models under our BAAT is also on par with that of the one trained on the benign dataset. An interesting phenomenon is that the BAs of our method are even larger than those of the cases under no attack. It is most probably because the style transfer used in our attack serves as an effective data augmentation to some extent (since we do not re-assign the label of poisoned samples), which is harmless or even beneficial. We will further explore it in our future work. These results verify the effectiveness of our attribute-based trigger patterns.

\subsection{Ablation Study}
In this section, we discuss the effects of key hyper-parameters involved in our BAAT. We adopt ResNet-18 as an example for discussions. Unless otherwise specified, all settings are the same as those illustrated in Section \ref{sec:settings}.

\subsubsection{The Effects of Trigger Pattern}

\begin{figure}[!t]
\centering
\subfigure[]{
\includegraphics[width=0.21\textwidth]{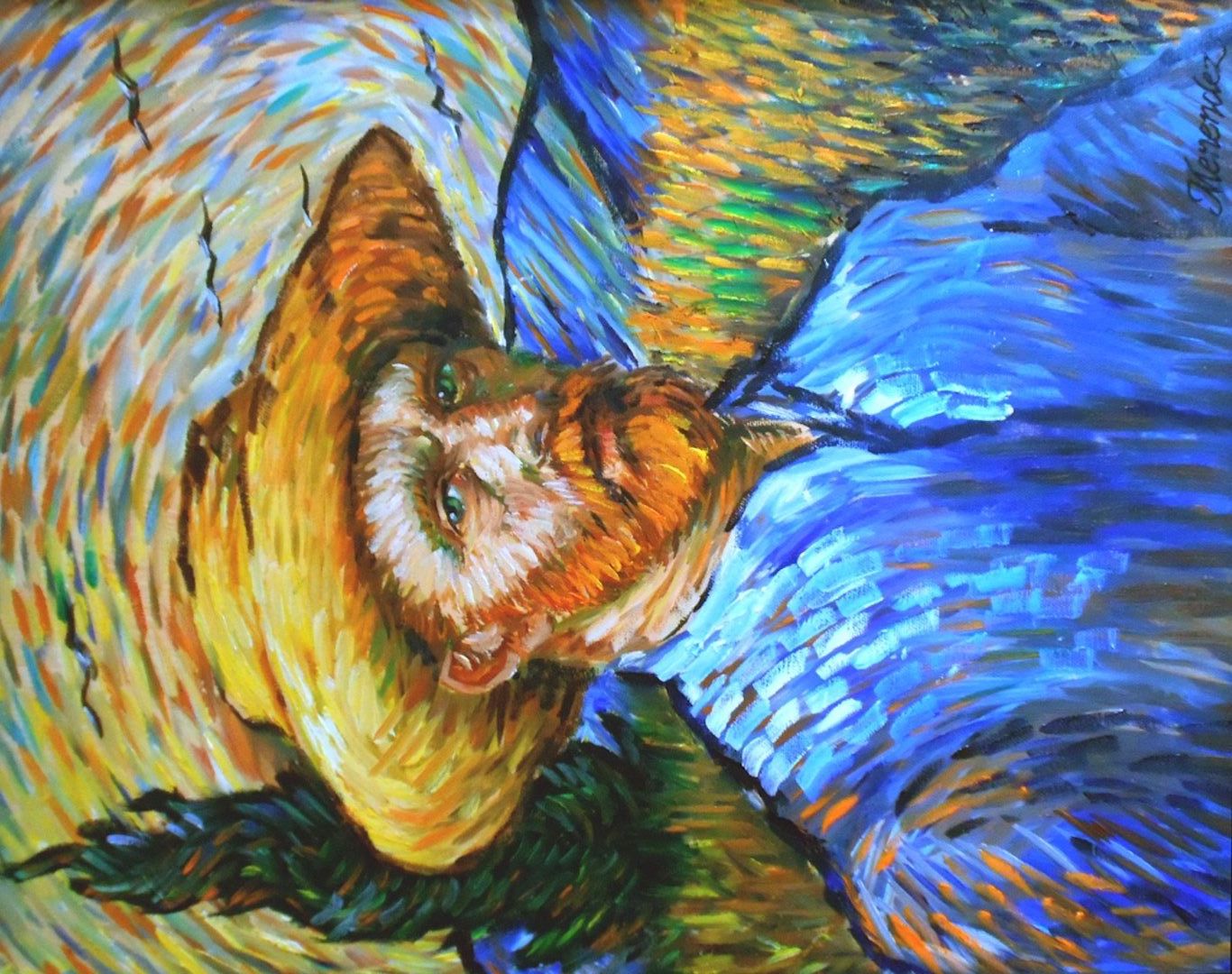}
}
\subfigure[]{
\includegraphics[width=0.22\textwidth]{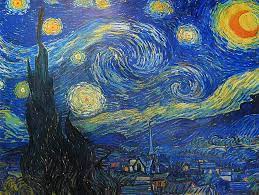}
}
\subfigure[]{
\includegraphics[width=0.22\textwidth]{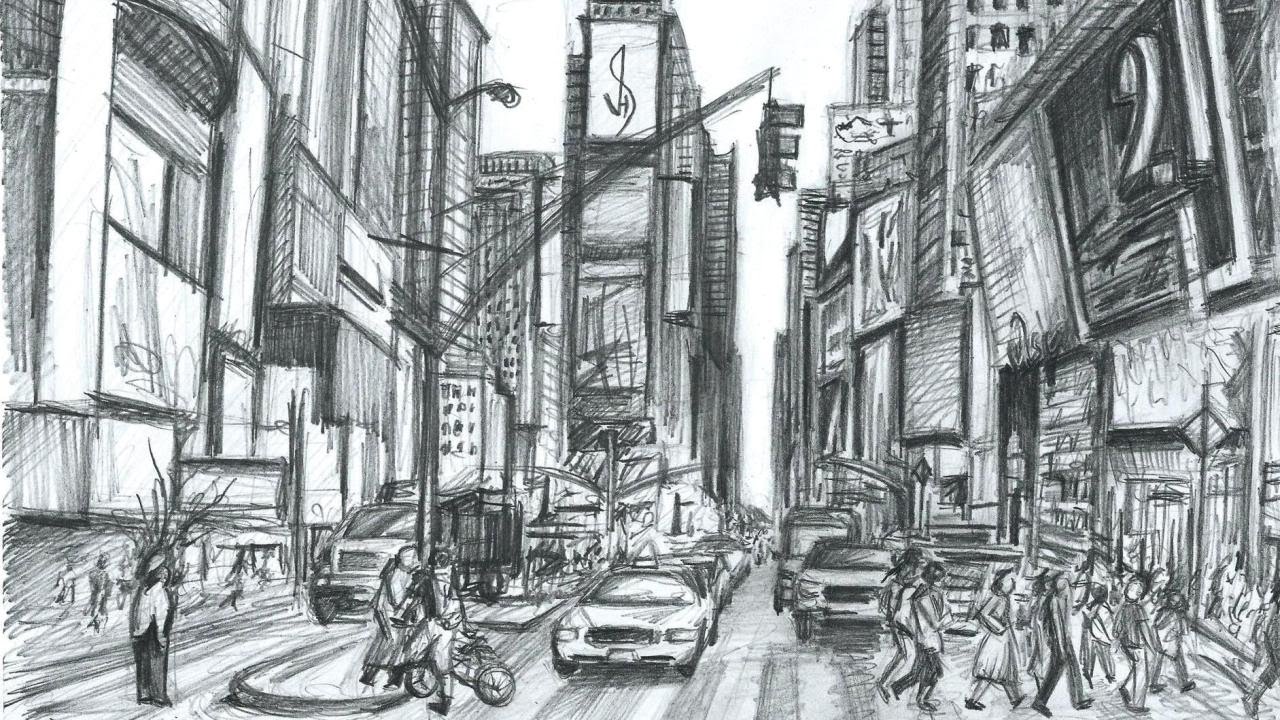}}
\subfigure[]{
\includegraphics[width=0.2\textwidth]{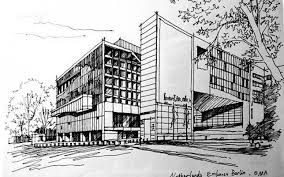}
}
\vspace{-0.5em}
\caption{Four style images used in our ablation study.}
\label{fig:styles}
\end{figure}

\noindent \textbf{Settings.} In this part, we discuss whether our method is still effective when using different trigger patterns. Specifically, we exploited four different hair types, including \textbf{a)} hi-top hairstyle with purple color, \textbf{b)} hi-top hairstyle with green color, \textbf{c)} jewrfro hairstyle with purple color, and \textbf{d)} jewrfro hairstyle with green color on the VGGFace2 dataset. Besides, we adopt four different style images (as shown in Figure \ref{fig:styles}) on the ImageNet dataset for discussions.

\begin{table}[!t]
\vspace{-0.5em}
\centering
\caption{The effectiveness of our BAAT method with different trigger patterns on VGGFace2 and ImageNet.}
\vspace{-0.8em}
\scalebox{1}{
\begin{tabular}{c|c|cccc}
\toprule
Dataset$\downarrow$                   & \tabincell{c}{Pattern$\rightarrow$\\Metric$\downarrow$} & (a)     & (b)     & (c)     & (d)     \\ \hline
\multirow{2}{*}{VGGFace2}  & BA (\%)         & 77.15 & 76.90 & 77.00 & 76.90 \\
                          & ASR (\%)        & 80.60 & 86.60 & 74.05 & 81.55 \\ \hline
\multirow{2}{*}{ImageNet} & BA (\%)         & 82.46 & 82.48 & 82.26 & 82.26 \\
                          & ASR (\%)        & 59.28 & 59.12 & 55.76 & 64.26 \\ \bottomrule
\end{tabular}
}
\label{tab:effects_pattern}
\vspace{-1.0em}
\end{table}

\noindent \textbf{Results.} As shown in Table \ref{tab:effects_pattern}, our BAAT is effective with each trigger pattern, although the performance may have some fluctuations. Specifically, the ASRs are larger than 70\% in all cases on the VGGFace2 dataset. These results verify that our BAAT method can reach promising attack performance with arbitrary adversary-specified trigger patterns.



\subsubsection{The Effects of Target Label}

To verify that our BAAT is still effective when different target labels are used, we evaluate our BAAT with four different labels. As shown in Table \ref{tab:effects_target}, our BAAT is effective in all cases, although the performance may have some fluctuations. For example, the ASRs are larger than 75\% in all cases on the VGGFace2 dataset. The ASRs are also larger than $55\%$ in all cases on the ImageNet dataset. These results verify the effectiveness of BAAT again.

\begin{table}[!t]
\centering
\caption{The effectiveness of our BAAT method with different target labels on VGGFace2 and ImageNet.}
\vspace{-0.8em}
\scalebox{1}{
\begin{tabular}{c|c|cccc}
\toprule
Dataset$\downarrow$                   & \tabincell{c}{Label$\rightarrow$\\Metric$\downarrow$} & 1     & 2     & 3     & 4     \\ \hline
\multirow{2}{*}{VGGFace2}  & BA (\%)         & 77.15 & 76.45 & 76.55 & 77.30 \\
                          & ASR (\%)        & 80.60 & 78.10 & 88.80 & 84.45 \\ \hline
\multirow{2}{*}{ImageNet} & BA (\%)         & 82.46 & 82.54 & 82.52 & 82.56 \\
                          & ASR (\%)        & 59.28 & 58.32 & 59.34 & 57.70 \\ \bottomrule
\end{tabular}
}
\vspace{-1.0em}
\label{tab:effects_target}
\end{table}




\subsubsection{The Effects of Poisoning Rate}

\begin{figure}[!t]
\centering
\subfigure[VGGFace2]{
\includegraphics[width=0.235\textwidth]{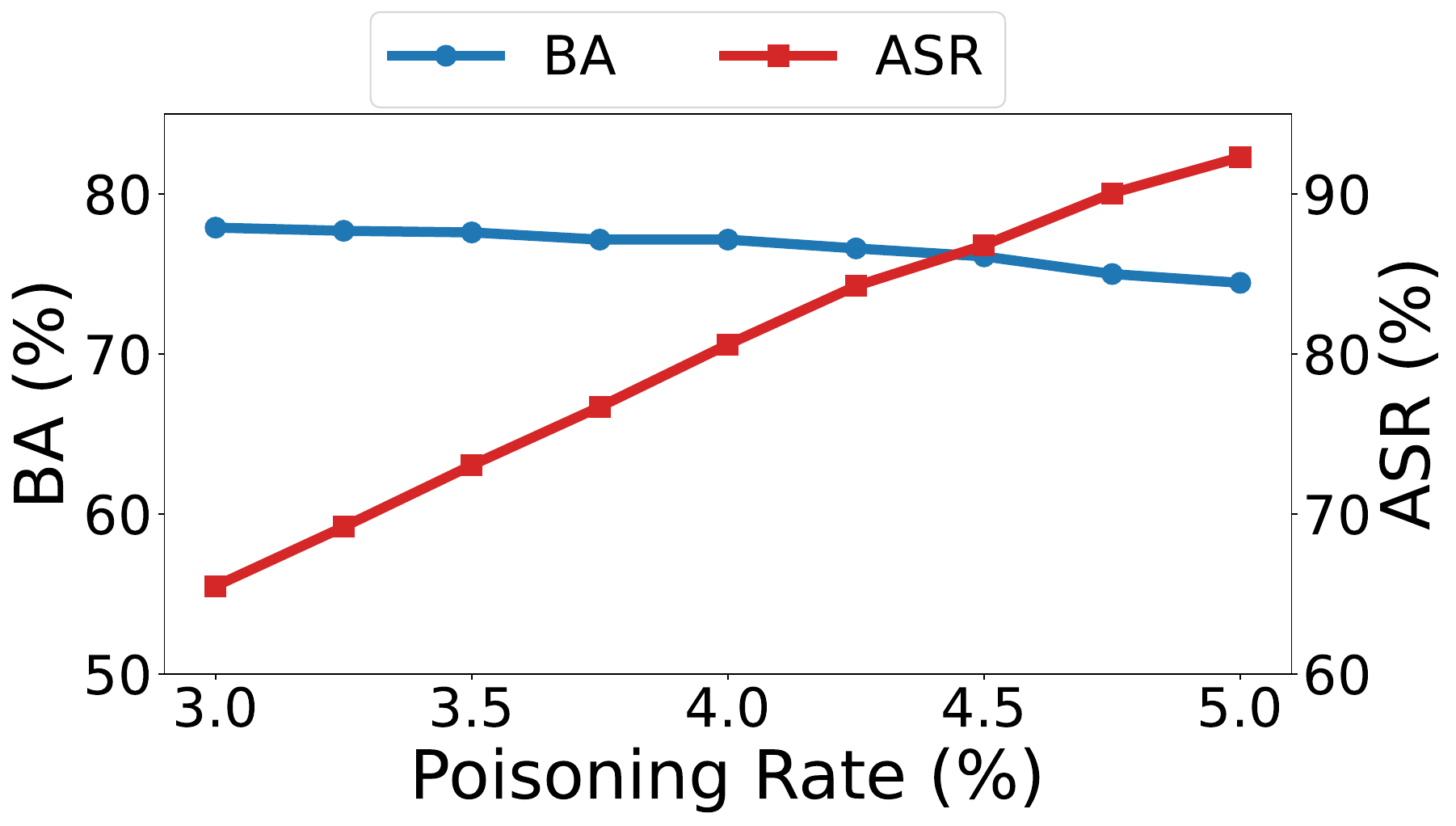}}
\subfigure[ImageNet]{
\includegraphics[width=0.235\textwidth]{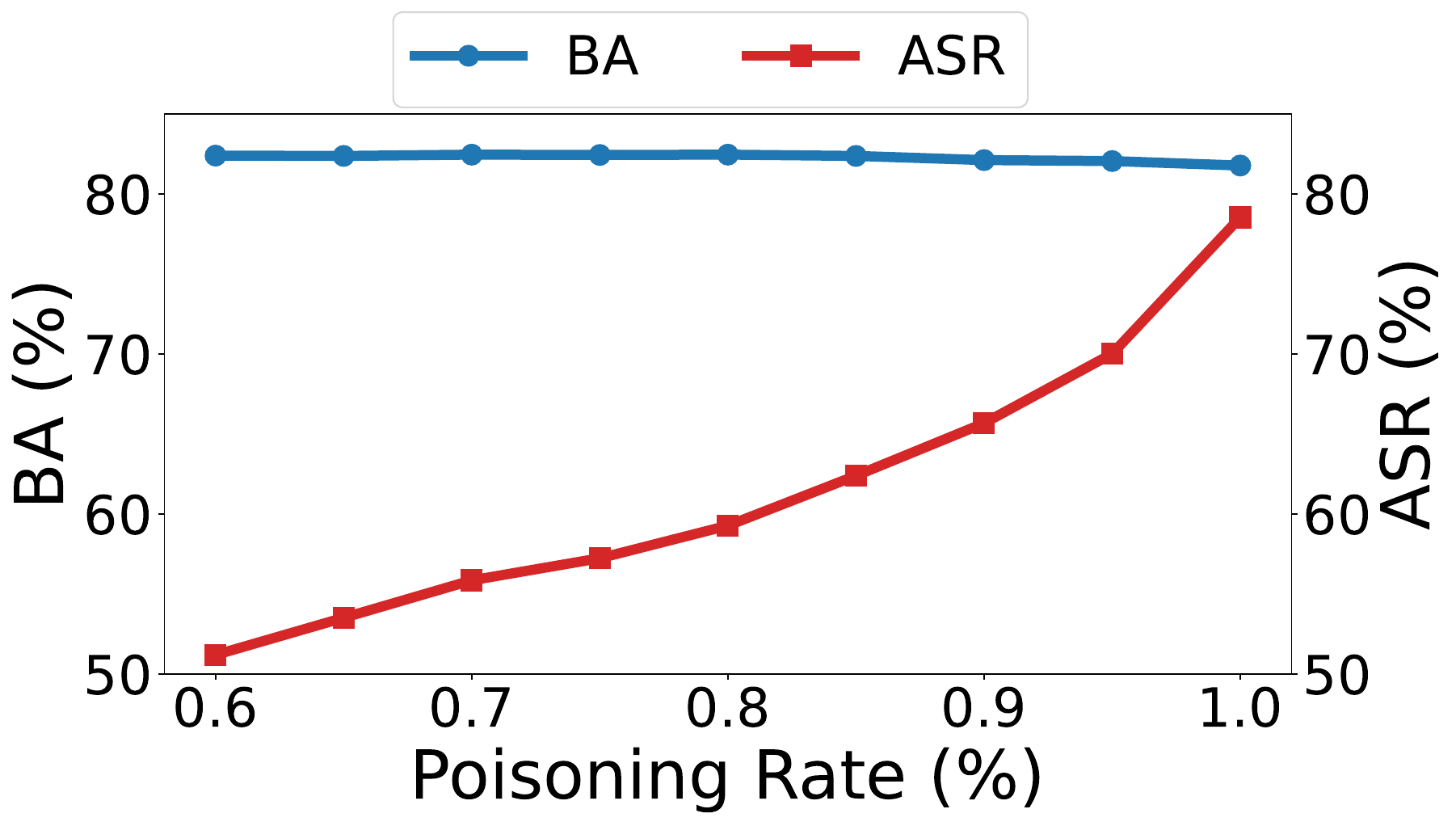}}
\vspace{-0.8em}
\caption{The effects of poisoning rate towards our BAAT on the VGGFace2 and the ImageNet dataset.}
\label{fig:effects_poisoning_rate}
\end{figure}

In this part, we analyze how the poisoning rate affects our BAAT. As shown in Figure \ref{fig:effects_poisoning_rate}, the attack success rate (ASR) increases with the increase of the poisoning rate $\gamma$. In particular, our BAAT reaches a high ASR ($>50\%$) on both datasets by poisoning only 60\% training samples from the target class (\ie, $\gamma=3\%$ on VGGFace2 and $\gamma=0.6\%$ on ImageNet). Besides, the benign accuracy (BA) decreases with the increase of $\gamma$, although the decline rate is relatively slow. In other words, there is a trade-off between ASR and BA to some extent. Accordingly, the adversaries should assign $\gamma$ based on their specific needs.

\subsection{The Resistance to Potential Defenses}
In this section, we verify that our BAAT is resistant to representative backdoor defenses. For simplicity, we hereby also adopt ResNet-18 for our discussions.

\subsubsection{The Resistance to Classical Model Repairing}

Model repairing intends to directly remove backdoors from the attacked models by modifying their parameters. In this part, we explore the resistance of our BAAT to two classical and representative methods, including fine-tuning \cite{liu2017neural,liu2018fine,yang2023backdoor} and model pruning \cite{liu2018fine,wu2021adversarial,zheng2022data}.

\vspace{0.3em}
\noindent \textbf{Settings.} For fine-tuning, we fine-tune the fully-connected layers of the attacked model with 50\% benign training samples 30 epochs and set the learning rate as 0.1. The benign accuracy and attack success rate is evaluated after each epoch; For model pruning, we conduct channel pruning \cite{he2017channel} on the output of the last convolutional layer with 10\% benign training samples on both datasets. The pruning rate is set to $\beta \in \{0\%, 2\%, \cdots, 98\%\}$.

\begin{figure}[!t]
\centering
\vspace{-0.5em}
\subfigure[VGGFace2]{
\includegraphics[width=0.235\textwidth]{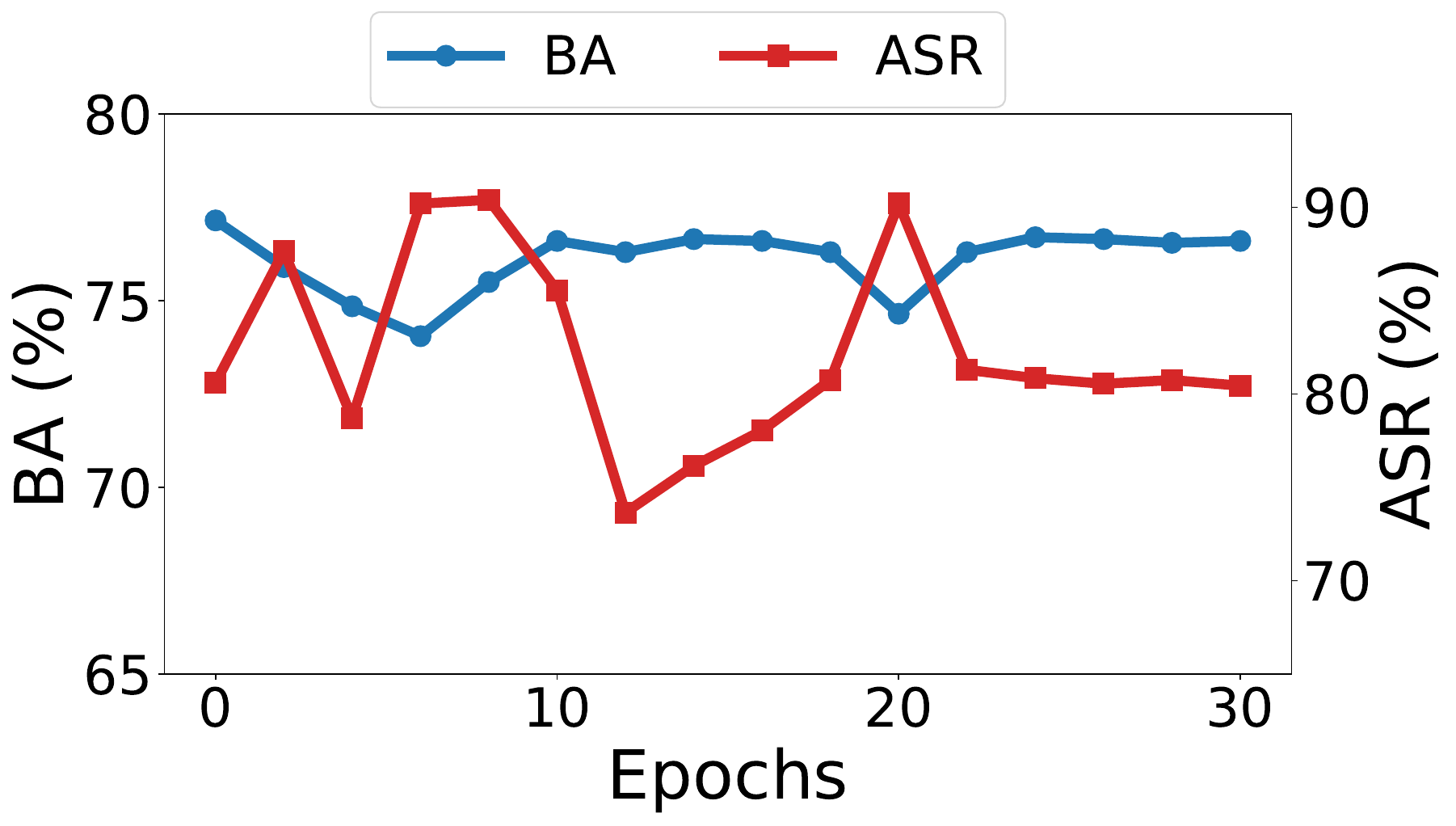}}
\subfigure[ImageNet]{
\includegraphics[width=0.235\textwidth]{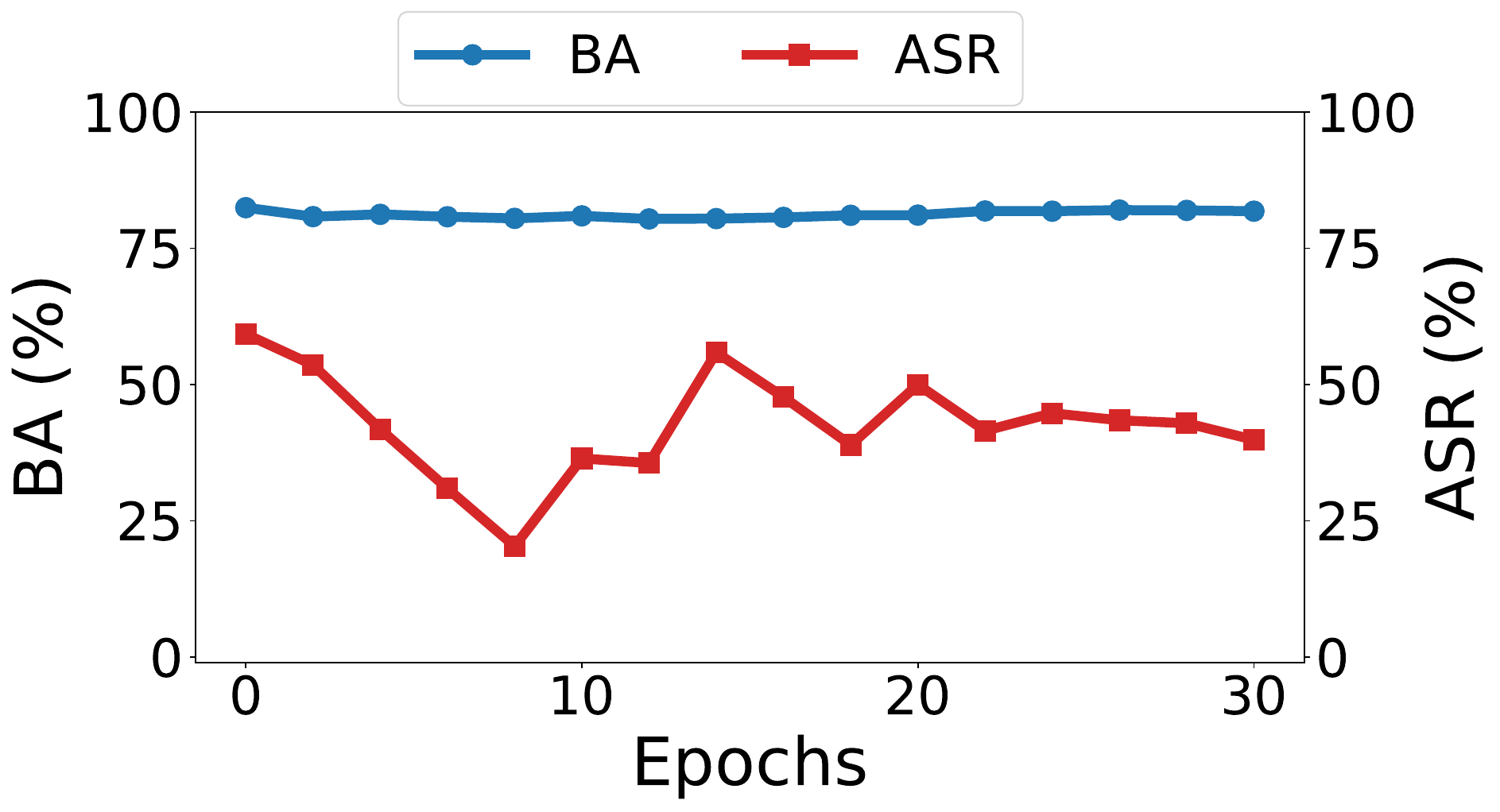}}
\vspace{-0.8em}
\caption{The resistance to fine-tuning.}
\label{fig:fine-tuning}
\vspace{-1em}
\end{figure}

\begin{figure}[!t]
\centering
\subfigure[VGGFace2]{
\includegraphics[width=0.235\textwidth]{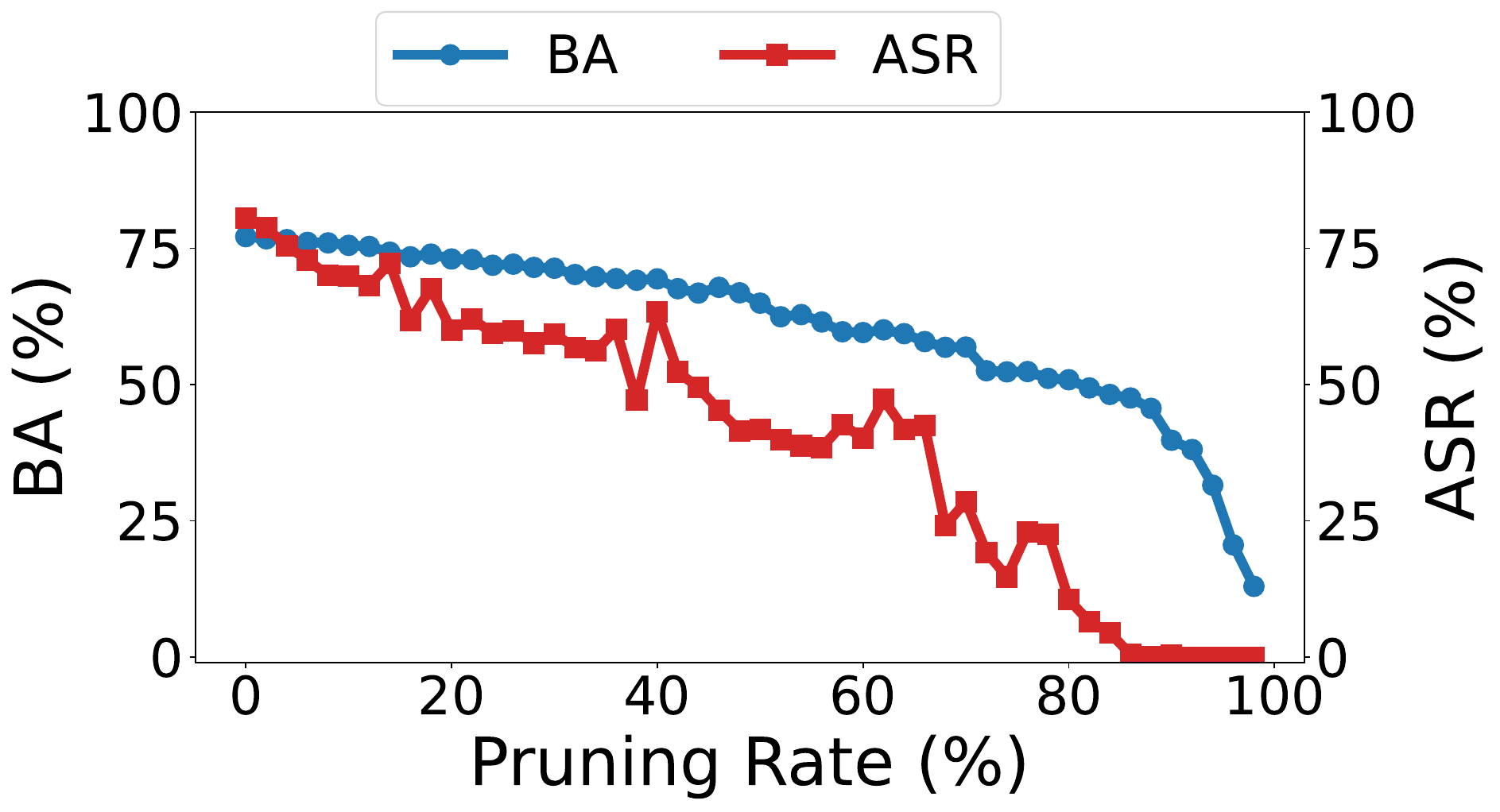}}
\subfigure[ImageNet]{
\includegraphics[width=0.235\textwidth]{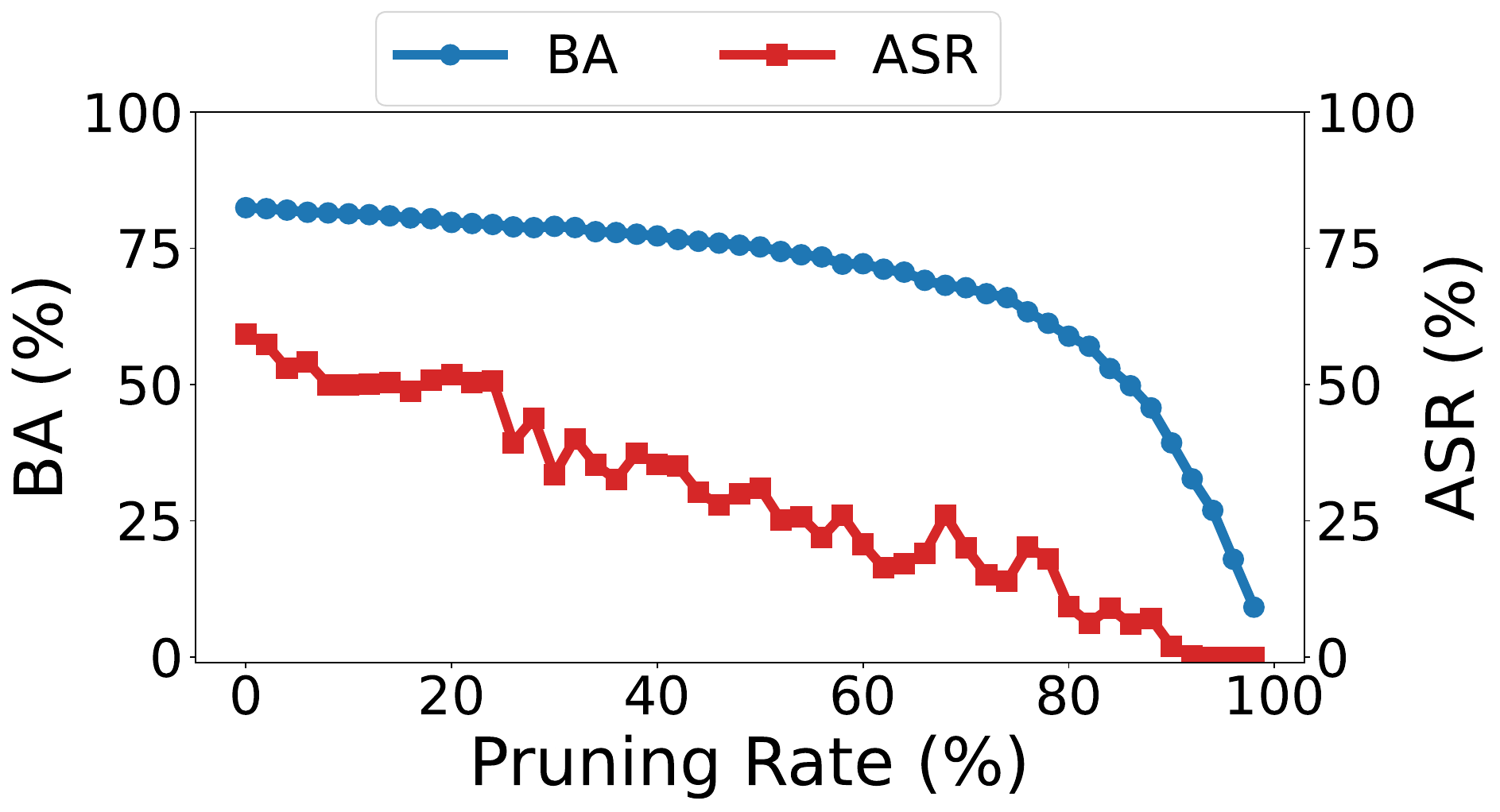}}
\vspace{-0.8em}
\caption{The resistance to model pruning.}
\label{fig:model-pruning}
\vspace{-0.3em}
\end{figure}

\vspace{0.3em}
\noindent \textbf{Results.} As shown in Figure \ref{fig:fine-tuning}-\ref{fig:model-pruning}, our method is resistant to fine-tuning and model pruning on both VGGFace2 and ImageNet datasets. Specifically, the attack success rate (ASR) is still larger than 70\% during the fine-tuning process on VGGFace2. Besides, model pruning can significantly reduce our ASR whereas with a great sacrifice of benign accuracy. These results verify the robustness of our BAAT method.

\begin{figure*}[!ht]
\centering
\subfigure[]{
\includegraphics[width=0.14\textwidth,frame]{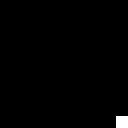}}\hspace{2em}
\subfigure[]{
\includegraphics[width=0.14\textwidth,frame]{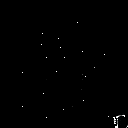}}
\hspace{2em}
\subfigure[]{
\includegraphics[width=0.14\textwidth,frame]{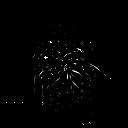}}
\hspace{2em}
\subfigure[]{
\includegraphics[width=0.14\textwidth,frame]{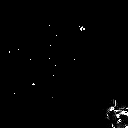}}
\hspace{2em}
\subfigure[]{
\includegraphics[width=0.14\textwidth,frame]{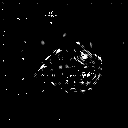}}

\vspace{-1em}
\caption{The ground-truth trigger pattern of BadNets and synthesized patterns of BadNets and our BAAT. \textbf{(a)} The ground-truth trigger pattern; \textbf{(b)\&(d)} The synthesized trigger patterns of BadNets on VGGFace2 and ImageNet, respectively; \textbf{(c)\&(e)} The synthesized trigger patterns of our BAAT on VGGFace2 and ImageNet, respectively.}
\label{fig:NC}
\vspace{-1em}
\end{figure*}

\begin{figure*}[!t]
\centering
\subfigure[VGGFace2]{
\includegraphics[width=0.46\textwidth]{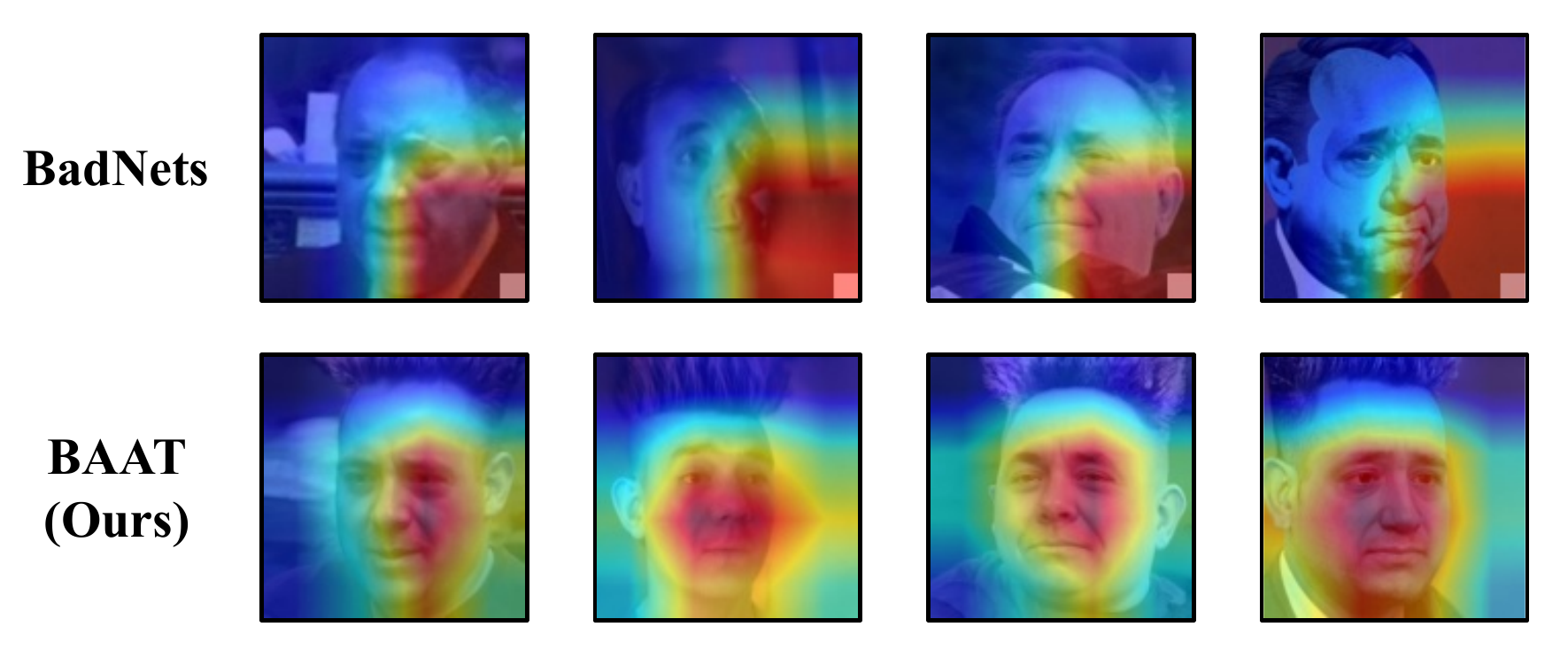}}
\hspace{2.5em}
\subfigure[ImageNet]{
\includegraphics[width=0.46\textwidth]{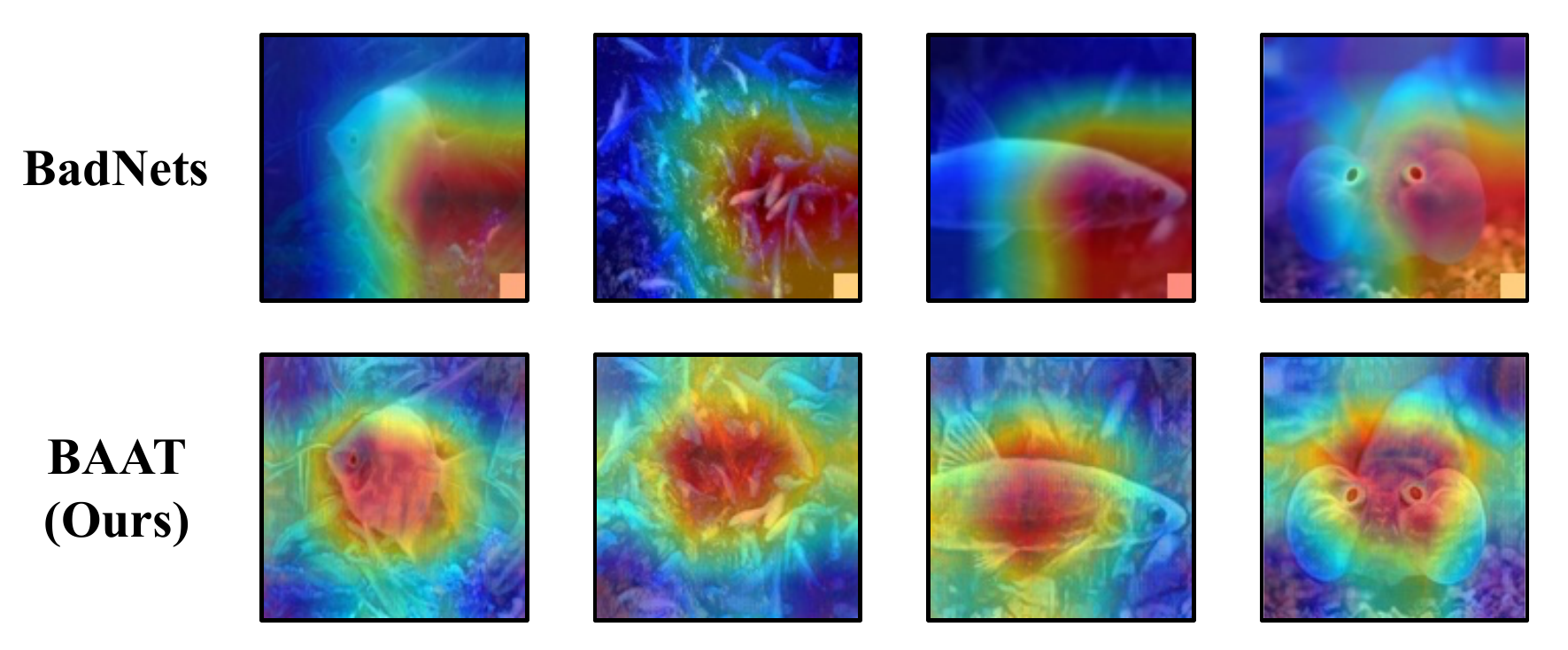}}
\vspace{-0.8em}
\caption{The Grad-CAM of poisoned samples generated by BadNets and our BAAT.}
\label{fig:Saliency_all}
\vspace{-0.5em}
\end{figure*}


\subsubsection{The Resistance to Advanced Model Repairing}

\noindent \textbf{Settings.} We hereby evaluate the resistance of our BAAT to advanced and representative model-repairing-based methods, including mode connectivity repair (MCR) \cite{zhao2020bridging} and neural attention distillation (NAD) \cite{li2021neural}. Specifically, for MCR, we adopt the model after fine-tuning as another attacked DNN and train a Bezier-type connect curve with 10\% benign training samples for 100 epochs. Besides, we set $t=0.2$ for repairing; For NAD, we set the hyper-parameter for the attention loss to 1. We implement both methods based on the codes provided in \texttt{BackdoorBox} \cite{li2023backdoorbox}.

\begin{table}[!t]
\centering
\caption{The resistance to MCR and NAD.}
\vspace{-0.8em}
\begin{tabular}{c|cc|cc}
\toprule
Dataset$\rightarrow$        & \multicolumn{2}{c|}{VGGFace2} & \multicolumn{2}{c}{ImageNet} \\ \hline
Method$\downarrow$, Metric$\rightarrow$ & BA            & ASR           & BA            & ASR          \\ \hline
No Defense     & 77.15         & 80.60         & 82.46         & 59.28        \\ \hline
MCR            & 77.65         & 17.60         & 82.06         & 43.08        \\
NAD            & 74.40         & 76.25         & 68.16         & 14.38        \\ \bottomrule
\end{tabular}
\vspace{-1.0em}
\label{tab:resistance_advancedMRD}
\end{table}

\vspace{0.3em}
\noindent \textbf{Results.} As shown in Table \ref{tab:resistance_advancedMRD}, our BAAT preserves a relatively high attack success rate ($>15\%$) after defenses in many cases. In particular, the ASR is still larger than 10\% on the ImageNet dataset under NAD, although it decreases the benign accuracy by nearly 15\%. In conclusion, our BAAT is also resistant to them to a large extent.

\subsubsection{The Resistance to Trigger-synthesis-based Defenses}
\label{sec:res_trigger-synthesis}

In this part, we show that our BAAT is also resistant to neural cleanse \cite{wang2019neural} and SentiNet \cite{chou2020sentinet}, which are two representative types of trigger-synthesis-based defenses.

\begin{table}[!t]
\centering
\caption{The resistance to Auto-Encoder and ShrinkPad.}
\vspace{-0.8em}
\begin{tabular}{c|cc|cc}
\toprule
Dataset$\rightarrow$        & \multicolumn{2}{c|}{VGGFace2} & \multicolumn{2}{c}{ImageNet} \\ \hline
Method$\downarrow$, Metric$\rightarrow$ & BA            & ASR           & BA            & ASR          \\ \hline
No Defense     & 77.15         & 80.60         & 82.46         & 59.28        \\ \hline
Auto-Encoder            & 73.85         & 68.55         & 64.74         & 47.20        \\
ShrinkPad            & 67.60         & 35.65         & 73.88         & 37.62        \\ \bottomrule
\end{tabular}
\vspace{-1.0em}
\label{tab:resistance_preprocessing}
\end{table}

\vspace{0.3em}
\noindent \textbf{Settings.} We adopt BadNets with a $12 \times 12$ white square located at the right corner of images for reference since it can be detected by neural cleanse and SentiNet. All other settings are the same as those presented in Section \ref{sec:settings}. For neural cleanse, we implement it based on its open-sourced codes and default settings; For SentiNet, we generate the saliency maps of DNNs attacked by BadNets and our BAAT, based on Grad-CAM \cite{selvaraju2017grad} with its default settings.

\vspace{0.3em}
\noindent \textbf{Results.} As shown in Figure \ref{fig:NC}, the synthesized pattern of BadNets is similar to their ground-truth trigger pattern, whereas that of our attack is meaningless (\ie, neither scattered throughout the whole image nor concentrated in the hair location.). Besides, as shown in Figure \ref{fig:Saliency_all}, SentiNet can distinguish trigger regions from those generated by BadNets, while it fails to detect those generated by our BAAT since it will focus on nearly the object outline or even the whole image. These results indicate that our attack resists both neural cleanse and SentiNet.



\subsubsection{The Resistance to Pre-processing-based Defenses}

In this part, we discuss whether our BAAT is resistant to auto-encoder-based pre-processing (dubbed `Auto-Encoder') \cite{liu2017neural} and ShrinkPad \cite{li2021backdoor}, which are two representative pre-processing-based defenses.

\vspace{0.3em}
\noindent \textbf{Settings.} We adopt a pre-trained auto-encoder trained on the ImageNet dataset for Auto-Encoder. Specifically, we first resize the images from $3 \times 128 \times 128$ to $3 \times 224 \times 224$ before feeding into the auto-encoder. After that, we shrink the pre-processed images back to $3 \times 128 \times 128$, based on which to calculate the benign accuracy and the attack success rate; We implement ShrinkPad based on \texttt{BackdoorBox} \cite{li2023backdoorbox}, where the shrinking size is set to 12 pixels on both datasets.

\vspace{0.3em}
\noindent \textbf{Results.} As shown in Table \ref{tab:resistance_preprocessing}, Auto-Encoder has minor benefits in reducing our attack success rate. The attack success rates are still larger than 45\% after Auto-Encoder on both datasets. It is mostly because our triggers are not additive perturbations with small magnitude, although they are still stealthy for human inspection. Besides, our attack is also resistant to ShrinkPad to a large extent, although it can decrease our ASR to some extent. It is mostly because our trigger patterns are large and not static.


\subsubsection{The Resistance to Sample-filtering-based Defenses}
In this part, we examine whether our attack can circumvent representative sample-level backdoor detection methods, including STRIP \cite{gao2022design} and SCALE-UP \cite{guo2023scaleup}.

\vspace{0.3em}
\noindent \textbf{Settings.} We adopt the same BadNets obtained in Section \ref{sec:res_trigger-synthesis} for comparative experiments on STRIP. Following the settings in \cite{guo2023scaleup}, we exploit a $12 \times 12$ random noise as a trigger pattern to train a new BadNets for comparative experiments on SCALE-UP. We implement STRIP and SCALE-UP based on their open-sourced codes.

\begin{table}[!t]
   \centering
   \caption{The entropy generated by STRIP of different attacks. The higher the entropy, the harder the detection.}
   \vspace{-0.8em}
   \begin{tabular}{cc|cc}
   \toprule
   \multicolumn{2}{c|}{VGGFace2} & \multicolumn{2}{c}{ImageNet} \\ \hline
   BadNets        & BAAT (Ours)        & BadNets        & BAAT (Ours)        \\
   0.220          & 0.814       & 0.446          & 1.039       \\ \bottomrule
   \end{tabular}
   \label{tab:resistance_STRIP}
   \vspace{-0.5em}
   \end{table}

\begin{table}[!t]
   \centering
   \caption{The AUROC of SCALE-UP in detecting BadNets and our BAAT on VGGFace2 and ImageNet datasets.}
   \vspace{-0.8em}
   \begin{tabular}{cc|cc}
   \toprule
   \multicolumn{2}{c|}{VGGFace2} & \multicolumn{2}{c}{ImageNet} \\ \hline
   BadNets        & BAAT (Ours)        & BadNets        & BAAT (Ours)        \\
   0.853          & 0.472       & 0.936          & 0.310       \\ \bottomrule
   \end{tabular}
   \label{tab:resistance_SCALE-UP}
   \vspace{-1em}
   \end{table}

\vspace{0.3em}
\noindent \textbf{Results.} As shown in Table \ref{tab:resistance_STRIP}, the entropy of our BAAT is significantly higher than that of BadNets on both datasets. These results indicate that STRIP can hardly detect our attack. Besides, as shown in Table \ref{tab:resistance_SCALE-UP}, our attack can also circumvent the detection of SCALE-UP, whereas BadNets cannot. These results verify the stealthiness of our BAAT.


\section{Discussions}

\subsection{The Analysis of Computational Complexity}
In general, BAAT introduces only a small overhead during the one-time trigger generation phase, but it maintains comparable training and inference efficiency to clean models. We analyze the efficiency of our method as follows.

\vspace{0.3em}
\noindent \textbf{Attack Phase.} In this phase, we employ an attribute editor to modify attributes and generate poisoned samples. Let $N$ and $\lambda$ represent the size of the benign dataset and the poisoning rate, respectively. The computational complexity of our method is $O(\lambda \cdot N)$, as BAAT only requires poisoning a small number of randomly selected samples during this phase. While this introduces additional computation, it is a one-time operation performed prior to training. For instance, on the ImageNet dataset, our method generates poisoned samples for $224\times224$ images at an average speed of 56.7 ms per image on an NVIDIA GeForce RTX 3080 Ti.

\vspace{0.3em}
\noindent \textbf{Training Phase.} Incorporating poisoned samples either not change the training process nor increase the number of training samples. As such, the computational cost of this process remains identical to that of the standard one.

\vspace{0.3em}
\noindent \textbf{Inference Phase.} During the inference process, BAAT does not require additional operations (\eg, data augmentation or post-processing) of the attacked model. As a result, the inference time is the same as that of a clean model. Note that the adversary needs to generate the attacked image locally through the same pre-trained attribute editor used for poisoning, but its cost is negligible (\ie, $O(1)$).

\subsection{The Comparison to Related Works}

\subsubsection{The Comparison to Data Poisoning}
As introduced in \cite{li2022backdoor}, there are two types of data poisoning, including classical data poisoning \cite{xiao2015feature} and advanced data poisoning \cite{schwarzschild2021just}. Specifically, the former intends to reduce model generalization, leading the attacked models to correctly predict training samples whereas having limited performance in predicting testing samples. The latter leads attacked models to have satisfied test accuracy while misclassifying some adversary-specified (unmodified) samples. Both our BAAT and data poisoning intend to implant malicious prediction behaviors by poisoning some training samples. However, they still have many intrinsic differences.

\vspace{0.3em}
\subsec{The Comparison to Classical Data Poisoning.} Firstly, our BAAT has a different purpose. Our attack preserves high accuracy in predicting benign testing samples while classical data poisoning is not. Accordingly, our method is more stealthy, since users can easily detect classical data poisoning by evaluating model performance on a local verification set while it has limited benefits in detecting our BAAT; Secondly, our method has a different mechanism. Specifically, the effectiveness of classical data poisoning is mostly due to the sensitiveness of the training process, so that even a small domain shift of training samples may lead to significantly different decision surfaces of attacked models. In contrast, BAAT relies on the data-driven model training process and domain shift between training and testing samples.

\vspace{0.3em}
\subsec{The Comparison to Advanced Data Poisoning.} Firstly, advanced data poisoning can only misclassify a few pre-defined images whereas our BAAT can lead to the misjudgments of all images containing the trigger pattern. It is mostly due to their second difference that the advanced data poisoning does not require modifying the images before feeding into attacked DNNs in the inference process. Thirdly, the effectiveness of advanced data poisoning is mainly because DNNs are over-parameterized and therefore the decision surface can have sophisticated structures near the adversary-specified samples for misclassification. It is also different from that of our BAAT.

\subsubsection{The Comparison to Adversarial Attacks}
Both our BAAT and adversarial attacks \cite{he2023generating} intend to make the DNNs misclassify samples during the inference process by adding malicious perturbations. However, they still have many essential differences, as follows.

Firstly, the success of adversarial attacks is mostly due to the behavior differences between DNNs and humans, which is different from that of our attack. Secondly, the malicious perturbations are known ($i.e.$, non-optimized) by BAAT whereas adversarial attacks need to obtain them based on the optimization process. As such, adversarial attacks cannot be real-time in many cases, since the optimization requires querying the DNNs multiple times under either white-box or black-box settings. Lastly, our BAAT requires modifying the training samples without any additional requirements in the inference process, while adversarial attacks need to control the inference process to some extent.

\subsubsection{The Comparison to Style-based Attacks}
We notice that there are a few other works \cite{duan2020adversarial,cheng2021deep} also focused on attacking DNNs based on style transfer. In this part, we compare our BAAT to them.

\cite{duan2020adversarial} adopted style transfer to generate adversarial examples in both digital and physical-world scenarios. Similar to existing adversarial attacks, this method obtained (style-based) perturbations by optimization, which takes time. Besides, it was designed under the white-box setting where the adversary can obtain the source files of the target model. In contrast, our BAAT does not have these limitations.

\cite{cheng2021deep} also adopted style transfer to design the backdoor attack, which is closely related to our method. However, this attack needed to control the training process of attacked DNNs, whereas our BAAT only needs to poison a few training samples. Besides, this attack was designed under the poisoned-label setting, whereas our method is under the clean-label setting. These differences make our attack more practical and therefore more threatening.

Besides, we need to notice that we only adopt style transfer as an example to discuss how to generate attribute triggers towards natural images. Users may use other methods, based on their domain knowledge of the target task.

\subsection{Potential Negative Societal Impacts \& Limitations}

In this paper, our main goal is to design a simple yet effective tool to evaluate the backdoor robustness of existing DNN-based classifiers. However, we notice that our BAAT is resistant to existing backdoor defenses and could be used by the backdoor adversaries for malicious purposes. The adversaries may also design similar attacks against other tasks inspired by our research. Although an effective defense is yet to be developed, one may mitigate or even avoid this threat via only using fully-trusted training resources. Our next step is to design
principled and advanced defenses against BAAT-type backdoor attacks.

We notice that our method cannot optimize the attribute trigger due to its discontinuity and non-differentiability, although using handcrafted attributes (as our BAAT does) has already achieved a sufficiently high attack success rate. Our work is only the first step towards clean-label sample-specific backdoor attacks. We will discuss how to optimize attribute triggers in our future works. We will also discuss how to generalize our BAAT method to other modalities, such as audio and texts, in the future.

\section{Conclusion}
\label{sec:conclusion}
In this paper, we revisited the sample-specific backdoor attack (SSBA). We revealed that existing SSBAs are not sufficiently stealthy due to their poisoned-label nature, where users can discover anomalies if they check the image-label relationship. We found that extending existing methods to the clean-label attacks simply by poisoning samples only from the target class has minor effects and its failure reasons. Based on our analyses, in this paper, we designed the backdoor attack with attribute trigger (BAAT) inspired by the decision process of humans. Our BAAT is the first effective sample-specific backdoor attack with clean labels. It was also resistant to existing defenses to a large extent. We hope that our attack can serve as a strong baseline to facilitate the design of more robust and secure DNNs.

\section*{Acknowledgments}

This research is supported in part by the National Key Research and Development Program of China under Grant 2021YFB3100300 and the National Natural Science Foundation of China under Grants (62171248, 62441238, 62072395, and U20A20178). This work was also partly done when Yiming Li was a research intern at Ant Group. We also sincerely thank Mr. Chengxiao Luo from Tsinghua University for his implementation of some preliminary experiments on the VGGFace2 dataset, and Prof. Yong Jiang from Tsinghua University and Dr. Haiqin Weng from Ant group for their valuable suggestions on an early draft of this paper.


\bibliographystyle{IEEEtran}
\bibliography{references}

\begin{onecolumn}
\appendices

\setcounter{theorem}{0}
\setcounter{equation}{0}

\section{The Proof of Theorem 1}

\begin{theorem}\label{lem1}
Suppose the training dataset consists of $N_b$ benign samples $\{(\bm{x}_i, y_i)\}_{i=1}^{N_b}$ and $N_p$ poisoned samples $\{(\bm{x}_j', y_t)\}_{j=1}^{N_p}$, whose images are i.i.d. sampled from uniform distribution and belonging to $K$ classes. Assume that the DNN $f(\cdot;\bm{\theta})$ is a multivariate kernel regression $K(\cdot)$ and is trained via 
$
     \min_{\bm{\theta}} \sum_{i=1}^{N_{b}} \mathcal{L}(f(\bm{x}_{i};\bm{\theta}),y_{i}) + \sum_{j=1}^{N_p} \mathcal{L}(f(\bm{x}'_{j};\bm{\theta}),y_t),
$
while trigger patterns are additive perturbations. Let $f^{(a)}$ and $f^{(s)}$ denote models attacked by sample-agnostic and sample-specific attacks, which select the same benign samples for poisoning on the same dataset, respectively. For their expected predictive confidences over the target label $y_t$, we have:

\begin{equation}
    \mathbb{E}_{\hat{\bm{x}}}[f^{(a)}(\hat{\bm{x}})] - \mathbb{E}_{\widetilde{\bm{x}}}[f^{(s)}(\widetilde{\bm{x}})] \geq 0,
\end{equation}
where $\hat{\bm{x}}$ and $\widetilde{\bm{x}}$ are poisoned testing samples of sample-agnostic and sample-specific attacks, respectively.

\end{theorem}

\comment{
For a given model to perform backdoor attacks, we have that the relation on expected predictive confidence over the target label $y_t$ has the following relation for training on sample-specific samles $f^{(a)}(\hat{\bm{x}})$ and that for the sample-static samples $f^{(s)}(\widetilde{\bm{x}})$:

\begin{flalign}
   &\lim_{\mathcal{D}_s\to \hat{\mathcal{D}}_{s} ; \mathcal{D}_b\to \hat{\mathcal{D}}_{b}} \mathbb{E} [f^{(a)}(\hat{\bm{x}})] - \mathbb{E}[f^{(s)}(\widetilde{\bm{x}})] \\ 
   &\geq C\frac{K(\bm{x_t},\bm{x_i})(1-e^{-2\gamma\bm{\Delta t}^{T}\bm{\Delta x}})}{(\sum_{i=1}^{N_p}K(\bm{x^{'}_{t}},\bm{x^{'}_{i}})+\sum_{i=1}^{N_b}K(\bm{x^{'}_{t}},\bm{x_{i}}))(\sum_{i=1}^{N_p}K(\bm{x^{'}_{t}},\bm{\hat{x}^{'}_{i}})+\sum_{i=1}^{N_b}K(\bm{x^{'}_{t}},\bm{x_{i}}))},
\end{flalign}
without loss of generality, we unify $\bm{\hat{x}}$ and $\widetilde{\bm{x}}$ as $\bm{x_t}$ to get rid of impact for inputs. where $K(x_t,x)=e^{-\gamma||x_t-x||_{2}^{2}}$ with $\gamma >0$, $C >0$, $\bm{\Delta t} = [\bm{t} - \bm{t_i}] ^{C\times H \times W}$, $\bm{\Delta x} = [\bm{x_t} - \bm{x_i}] ^{C\times H \times W}$.
}

\begin{proof}

We have $\bm{x^{'}_{t}}=\bm{x_t}+\bm{t}$ for poisoned samples since trigger patterns are additive. As such, for sample-specific attacks, we have $\bm{\widetilde{x}^{'}_{i}}=\bm{x_i}+\bm{t_{i}}$, while for the sample-agnostic attacks:  $\bm{\hat{x}^{'}_{i}}=\bm{x_i}+\bm{t}$, where $t$ represents the backdoor trigger. 

We treat our model as a k-way kernel least square classifier and use a cross-entropy loss for training the kernel, and the output of $f(\cdot)$ is a k-dimensional vector. Let us assume $\phi_{t}(\cdot) \in \mathbb{R}$ be expected predictive confidences corresponding to the target class $t$. Following previous works~\cite{jacot2018neural,guo2022aeva}, we know the kernel regression solution is:
\begin{equation}
    \phi_t(\cdot)=\frac{\sum_{i=1}^{N_b}K(\cdot, \bm{x_i})\cdot \bm{y_{i}}+\sum_{i=1}^{N_p}K(\cdot, \bm{x'_{i}})\cdot \bm{y_{t}}}{\sum_{i=1}^{N_b}K(\cdot, \bm{x_i})+\sum_{i=1}^{N_p}K(\cdot, \bm{x'_{i}})},
\end{equation}
where $K$ is the RBF kernel, $\bm{y}$ is the one-hot version of the label $y$.

We assume the training samples are evenly distributed, thus there are $\frac{N_b}{k}$ benign samples belonging to $y_t$. Without loss of generality,  we here let the target label $y_{t}=1$ while others are 0. Then, the regression solution can be re-formulated as:
\begin{equation}
    \phi_t(\cdot)=\frac{\sum_{i=1}^{N_b/k}K(\cdot, \bm{x_i})+\sum_{i=1}^{N_p}K(\cdot, \bm{x'_{i}})}{\sum_{i=1}^{N_b}K(\cdot, \bm{x_i})+\sum_{i=1}^{N_p}K(\cdot, \bm{x'_{i}})}.\label{eq:kernel2}
\end{equation}

Accordingly, for sample-specific attacks and sample-agnostic attacks, we respectively have:

\begin{equation}
    \mathbb{E}_{\widetilde{\bm{x}}}[f^{(s)}(\widetilde{\bm{x}})] 
\triangleq \phi_t(\bm{x^{'}_{t}})=\frac{\sum_{i=1}^{N_b/k}K(\bm{x^{'}_{t}}, \bm{x_i})+\sum_{i=1}^{N_p}K(\bm{x^{'}_{t}}, \bm{\widetilde{x}^{'}_{i}})}{\sum_{i=1}^{N_b}K(\bm{x_{t}^{'}}, \bm{x_i})+\sum_{i=1}^{N_p}K(\bm{x_{t}^{'}}, \bm{\widetilde{x}^{'}_{i}})}, 
\mathbb{E}_{\hat{\bm{x}}}[f^{(a)}(\hat{\bm{x}})] \triangleq
    \phi_t(\bm{x^{'}_{t}})=\frac{\sum_{i=1}^{N_b/k}K(\bm{x^{'}_{t}}, \bm{x_i})+\sum_{i=1}^{N_p}K(\bm{x^{'}_{t}}, \bm{\hat{x}'_{i}})}{\sum_{i=1}^{N_b}K(\bm{x_{t}^{'}}, \bm{x_i})+\sum_{i=1}^{N_p}K(\bm{x_{t}^{'}}, \bm{\hat{x}'_{i}})}.
\label{eq:kernel3}
\end{equation}


Accordingly, we have 

\begin{flalign}
   & \mathbb{E}_{\hat{\bm{x}}}[f^{(a)}(\hat{\bm{x}})] - \mathbb{E}_{\widetilde{\bm{x}}}[f^{(s)}(\widetilde{\bm{x}})] \\ 
   &= \frac{(\sum_{i=1}^{N_p} K(\bm{x'_t},\bm{\widetilde{x}^{'}_{i}})-\sum_{i=1}^{N_p} K(\bm{x'_t},\bm{\hat{x}'_{i}}))\sum_{i=1}^{N_b/k}K(\bm{x^{'}_{t}},\bm{x_{i}})-(\sum_{i=1}^{N_p} K(\bm{x'_t},\bm{\widetilde{x}^{'}_{i}})-\sum_{i=1}^{N_p} K(\bm{x'_t},\bm{\hat{x}'_{i}}))\sum_{i=1}^{N_b}K(\bm{x^{'}_{t}},\bm{x_{i}})}{(\sum_{i=1}^{N_p}K(\bm{x^{'}_{t}},\bm{\widetilde{x}^{'}_{i}})+\sum_{i=1}^{N_b}K(\bm{x^{'}_{t}},\bm{x_{i}}))(\sum_{i=1}^{N_p}K(\bm{x^{'}_{t}},\bm{\hat{x}^{'}_{i}})+\sum_{i=1}^{N_b}K(\bm{x^{'}_{t}},\bm{x_{i}}))},\\
   &=C \cdot \frac{\sum_{i=1}^{N_p} K(\bm{x'_t},\bm{\hat{x}'_{i}})-\sum_{i=1}^{N_p} K(\bm{x'_t},\bm{\widetilde{x}^{'}_{i}})}{(\sum_{i=1}^{N_p}K(\bm{x^{'}_{t}},\bm{\widetilde{x}^{'}_{i}})+\sum_{i=1}^{N_b}K(\bm{x^{'}_{t}},\bm{x_{i}}))(\sum_{i=1}^{N_p}K(\bm{x^{'}_{t}},\bm{\hat{x}^{'}_{i}})+\sum_{i=1}^{N_b}K(\bm{x^{'}_{t}},\bm{x_{i}}))}, 
   \label{eq:diff}
\end{flalign}
where $C=\sum_{i=1}^{N_b}K(\bm{x^{'}_{t}},\bm{x_{i}})-\sum_{i=1}^{N_b/k}K(\bm{x^{'}_{t}},\bm{x_{i}})$. In particular, we know that $C > 0$ since $\{\bm{x}_{i}\}_{i=1}^{{N_b}/k}$ belongs to $\{\bm{x}_{i}\}_{i=1}^{{N_b}}$.

For the upper term in the above equation (\ref{eq:diff}), due to the property of RBF kernel, we have:
\begin{flalign}
    &\sum_{i=1}^{N_p} K(\bm{x'_t},\bm{\hat{x}'_{i}})-\sum_{i=1}^{N_p} K(\bm{x'_t},\bm{\widetilde{x}^{'}_{i}})
    =\sum_{i=1}^{N_p} e^{-\gamma||\bm{x'_t}-\bm{\hat{x}^{'}_{i}}||_2^2}-e^{-\gamma||\bm{x'_t}-\bm{\widetilde{x}^{'}_{i}}||_2^2} =\sum_{i=1}^{N_p} e^{-\gamma||\bm{x_t}+\bm{t}-\bm{x_i}-\bm{t}||_2^2}- e^{-\gamma||\bm{x_t}+\bm{t}-\bm{x_i}-\bm{t_i}||_2^2}\\
    &=\sum_{i=1}^{N_p} e^{-\gamma||\bm{x_t}-\bm{x_i}||_2^2} (1-e^{-\gamma||\bm{t}-\bm{t_{i}}||_2^2}\cdot e^{-2\gamma\bm{\Delta t}^{T}\bm{\Delta x}})
    \geq \sum_{i=1}^{N_p} e^{-\gamma||\bm{x_t}-\bm{x_i}||_2^2} (1-e^{-2\gamma\bm{\Delta t}^{T}\bm{\Delta x}}) \geq \sum_{i=1}^{N_p} K(\bm{x_t}, \bm{x_{i}})(1-e^{-2\gamma\bm{\Delta t}^{T}\bm{\Delta x}}), 
\end{flalign}


\noindent where $\bm{\Delta t} = [\bm{t} - \bm{t_i}] ^{C\times H \times W}$, $\bm{\Delta x} = [\bm{x_t} - \bm{x_i}] ^{C\times H \times W}$, and $\gamma > 0$.

Put all above together, we have:

\begin{equation}
    \mathbb{E}_{\hat{\bm{x}}}[f^{(a)}(\hat{\bm{x}})] - \mathbb{E}_{\widetilde{\bm{x}}}[f^{(s)}(\widetilde{\bm{x}})]
   \geq C \cdot \frac{K(\bm{x_t},\bm{x_i})(1-e^{-2\gamma\bm{\Delta t}^{T}\bm{\Delta x}})}{(\sum_{i=1}^{N_p}K(\bm{x^{'}_{t}},\bm{\widetilde{x}^{'}_{i}})+\sum_{i=1}^{N_b}K(\bm{x^{'}_{t}},\bm{x_{i}}))(\sum_{i=1}^{N_p}K(\bm{x^{'}_{t}},\bm{\hat{x}^{'}_{i}})+\sum_{i=1}^{N_b}K(\bm{x^{'}_{t}},\bm{x_{i}}))} \geq 0. 
\end{equation}

\end{proof}

\end{onecolumn}

\end{document}